\documentclass[11pt,a4paper]{article}
\usepackage{jheppub}

\usepackage{tabularx}
\usepackage{amsmath,amssymb}
\usepackage{appendix}
\usepackage{graphicx}
\usepackage{xcolor}
\usepackage{cancel}
\usepackage{slashed}
\usepackage{CJKutf8}
\usepackage{multirow}
\usepackage{changes}
\usepackage{booktabs}
\usepackage{hyperref}
\usepackage{bbding}
\usepackage{placeins}
\usepackage{afterpage}

\graphicspath{{figure/}}

\newcommand{\beq}{\begin{eqnarray}}
\newcommand{\eeq}{\end{eqnarray}}

\makeatletter
\newcommand{\Rmnum}[1]{\expandafter\@slowromancap\romannumeral #1@}
\makeatother

\title{Light Primordial Black Holes from a First-Order Phase Transition near the End of Inflation: Baryogenesis and Gravitational Wave Signatures}

\author[a]{Zihong Cheng}
\affiliation[a]{MOE Key Laboratory of TianQin Mission, TianQin Research Center for
Gravitational Physics \& School of Physics and Astronomy, Frontiers
Science Center for TianQin, Gravitational Wave Research Center of CNSA, 
Sun Yat-sen University (Zhuhai Campus), Zhuhai 519082, China}
\emailAdd{chengzh83@mail2.sysu.edu.cn}

\author[a]{Fa Peng Huang\footnote{Corresponding author.}}
\emailAdd{huangfp8@sysu.edu.cn}

\abstract{
We investigate the formation and cosmological consequences
of light primordial black holes (PBHs) from vacuum bubbles
nucleated during a spectator sector first-order phase transition
in inflation. We construct a phenomenological model in which
the spectator potential evolves with the rolling inflaton,
allowing vacuum bubbles to nucleate and grow during inflation.
The vacuum ordering subsequently reverses near the end of
inflation, leaving the bubble interiors in a metastable false
vacuum state. We examine their stability against reverse
vacuum tunneling and impose benchmark conditions to suppress
this process before and after reheating. Sufficiently large
bubbles can subsequently undergo supercritical evolution
and form PBHs. In this framework, the PBH abundance is
determined by the vacuum tunneling history, while the
characteristic mass depends on the nucleation epoch and
bubble growth during inflation. For sufficiently light PBHs
that evaporate before big bang nucleosynthesis, Hawking
radiation can create locally superheated regions in which
electroweak symmetry is restored, potentially generating
a baryon asymmetry in the presence of additional CP-violating
interactions. We also investigate the direct gravitational wave (GW) signals
from Hawking radiation and the induced GWs from PBH density
fluctuations, the latter arising if an early PBH dominated
epoch occurs. For representative benchmark parameters, the
resulting baryon asymmetry can reach the ten-percent level
of the observed value, while the induced
GW signal can reach a peak amplitude of
$\Omega_{\rm GW,0}h^2\sim10^{-8}$ at frequencies of order
$10$--$100\,{\rm Hz}$.

}


\begin{document}

\begin{CJK}{UTF8}{gbsn}

\maketitle

\section{Introduction}
\label{sec:introduction}

First-order phase transitions (FOPTs)~\cite{Mazumdar:2018dfl,Hindmarsh:2020hop,Athron:2023xlk,Wang:2020jrd}
in the early Universe can provide mechanisms for dark matter
production in different forms, including nonthermally produced
dark sector particles~\cite{Katz:2016adq,Mansour:2023fwj,Shakya:2023kjf,Giudice:2024tcp,
Ai:2024ikj,Cataldi:2025nac,Cheng:2026npt,Ghoshal:2026hev}
and primordial black holes (PBHs)~\cite{Khlopov:1998nm,Khlopov:1999ys,Jung:2021mku,Liu:2021svg,
Lewicki:2023ioy,Gouttenoire:2023naa,Lewicki:2024ghw,Kanemura:2024pae,
Murai:2025hse,Ning:2026nfs}.
PBHs are themselves well-motivated dark matter candidates
and can contribute to the dark matter abundance over
phenomenologically allowed mass ranges~\cite{Green:2020jor,Carr:2021bzv}.
Depending on the nature of the transition and the dynamics
of vacuum bubbles, PBH formation associated with FOPTs
can proceed through several different channels, including
bubble collisions~\cite{Khlopov:1998nm,Khlopov:1999ys,Jung:2021mku}
and the collapse of false vacuum regions~\cite{Liu:2021svg,Lewicki:2023ioy,Gouttenoire:2023naa,Ning:2026nfs}.

The possibility of forming wormholes and PBHs from vacuum bubbles
was explored in early studies of cosmological
FOPTs~\cite{Sato:1981bf,Kodama:1981gu,Kodama:1982sf}.
A distinct possibility, of particular relevance to the present work,
is PBH formation from vacuum bubbles nucleated during inflation,
as discussed in Ref.~\cite{Garriga:2012bc} and further developed
in subsequent studies~\cite{Garriga:2015fdk,Deng:2017uwc,
Deng:2020mds,Maeso:2021xvl,Huang:2023mwy,Wang:2025hwc}.
During inflation, these bubbles can grow through both cosmic
expansion and bubble wall propagation, potentially reaching
superhorizon scales. Depending on their size, interior vacuum
energy, and wall tension, their subsequent evolution can follow
either a subcritical or a supercritical
branch~\cite{Garriga:2015fdk,Deng:2017uwc,Deng:2020mds}.
Subcritical bubbles may collapse to form black holes, although
their evolution can be affected by shape fluctuations originating
from quantum fluctuations at nucleation, which may lead to
fragmentation~\cite{Deng:2017uwc}.
In contrast, the interior of a supercritical bubble continues
to inflate and becomes connected to the exterior through a
wormhole, with a black hole observed in the parent universe.
This branch is less sensitive to the shape fluctuations that
can affect subcritical collapse~\cite{Deng:2017uwc}.
We therefore focus on supercritical bubbles as a more robust
channel for PBH formation. 
A spectator sector provides a suitable setting for this
mechanism, as it can undergo nontrivial vacuum dynamics
during inflation while remaining subdominant in the
total energy density~\cite{Maeso:2021xvl,Cheng:2026npt,Luo:2025ewp}.
Inflaton-dependent vacuum tunneling has also been
investigated as a mechanism for generating a localized
PBH mass spectrum~\cite{Wang:2025hwc}.
In the present work, rather than considering PBHs as
dark matter candidates, we focus on the formation
of light PBHs and their cosmological implications,
particularly baryogenesis and  gravitational wave (GW)
signatures associated with their evaporation.

To realize this mechanism, we construct a phenomenological
model in which the rolling inflaton changes the effective
spectator mass and hence the ordering of the spectator vacua.
 Before the transition, the spectator field $s$ preferentially occupies the vacuum at $s=0$. Around $N\simeq N_\star$, where $N$ denotes the number
of $e$-folds before the end of inflation and $N_\star$
characterizes the nucleation epoch, the nonzero minimum
becomes lower in energy,
allowing Coleman--De~Luccia (CDL)
vacuum tunneling~\cite{Coleman:1980aw,Lee:2014uza} and the nucleation
of bubbles of the nonzero phase. These bubbles can subsequently expand
while the nonzero phase remains energetically favored. As inflation
proceeds toward its end, the vacuum ordering reverses, leaving the
bubble interiors in a positive energy metastable state.
 This introduces an important consistency requirement
for subsequent PBH formation: reverse vacuum tunneling inside
these bubbles must remain sufficiently suppressed, particularly
after reheating when the bubbles can still be superhorizon.
We take this requirement into account by imposing benchmark
stability conditions on the metastable interior before and
after reheating. Provided that these conditions are satisfied,
sufficiently large bubbles can undergo supercritical evolution
and form PBHs. In this framework, the PBH abundance is determined
by the vacuum tunneling history, while the characteristic mass
is controlled by the nucleation epoch and the bubble growth
history during inflation, with the accumulated wall propagation
parameterized by $v_{\rm eff}$. Both quantities are thus linked
to the underlying spectator sector dynamics rather than
introduced as independent phenomenological inputs.

The critical mass separating the subcritical and supercritical
regimes depends on the characteristic scales of the bubble
and can vary over a broad range as the underlying model
parameters are changed~\cite{Garriga:2015fdk}, leaving open
the possibility of forming sufficiently light PBHs in the
supercritical regime. The subsequent Hawking evaporation
of such PBHs~\cite{Hawking:1974rv,MacGibbon:1991vc,Auffinger:2022khh}
can have important consequences for the early Universe.
In particular, Hawking radiation can provide a nonthermal
source of dark matter particles~\cite{Cheek:2021odj,Morrison:2018xla,Jiang:2025blz,Kuwahara:2025eeo},
while entropy injection from PBH evaporation can dilute
a pre-existing baryon asymmetry~\cite{Calabrese:2023key}.
The energy deposited by Hawking radiation can also create
localized hot spots, potentially affecting dark matter
production and baryogenesis~\cite{Gunn:2024xaq}.
Independently of evaporation, pre-existing PBHs may modify
vacuum bubble nucleation during FOPTs
through their gravitational effects~\cite{Chang:2025rda}.

Of particular interest to the present work is baryogenesis
during the final explosive stage of PBH
evaporation~\cite{Klipfel:2026nzx}.
Energetic Hawking radiation deposits energy into the
surrounding plasma and can create a locally superheated
region in which electroweak symmetry is
restored~\cite{Vanvlasselaer:2026vkh}.
In the presence of additional CP-violating interactions,
a local baryon asymmetry can be generated at the moving
interfaces of these regions. If the PBH explosions occur
after sphaleron processes have become inefficient in the
ambient plasma, the generated asymmetry may survive
subsequent washout. The final cosmological baryon asymmetry
then depends on the PBH abundance, the entropy produced
during evaporation, and the thermal and hydrodynamic
conditions governing the final explosive stage.

The same PBH population can also generate
GWs through two physically distinct mechanisms. Gravitons
are emitted directly through Hawking radiation, producing a
stochastic GW background extending to very high
frequencies~\cite{Ireland:2023avg,Dolgov:2011cq}.
In addition, if the PBH population becomes dynamically
dominant before evaporation, fluctuations in the PBH number
density can source scalar perturbations and subsequently
induce GWs at second
order~\cite{Domenech:2020ssp,Domenech:2024wao,
Borah:2026zbl,Jiang:2025blz}.
The GW contribution from Hawking radiation is determined by the evaporation
history of the PBHs, whereas the induced GW contribution
depends strongly on the PBH abundance and on the existence
of an intermediate PBH dominated epoch.


\afterpage{%
    \par\noindent
    \makebox[\linewidth][c]{%
        \includegraphics[
            width=\textwidth,
            height=1\textheight,
            keepaspectratio
        ]{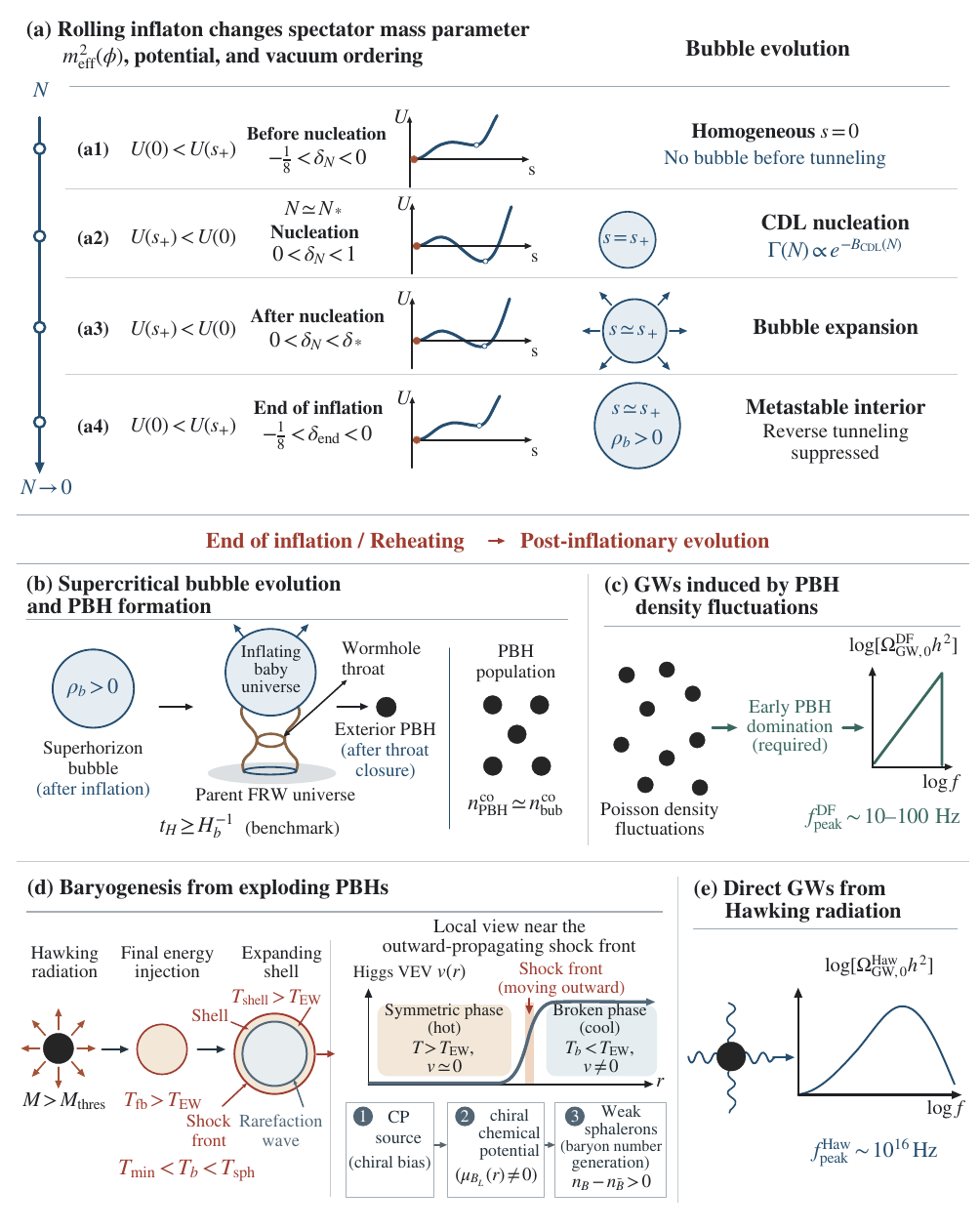}%
    }%
    \par\smallskip

    \refstepcounter{figure}
    \label{fig:PBH_mechanism}

    \begingroup
    \small
    \noindent\textbf{Figure~\thefigure.}\quad
    Schematic overview of the formation of light PBHs
    from a spectator sector FOPT during inflation,
    their subsequent Hawking evaporation,
    and the associated cosmological signatures.
    (a) Four representative stages of the inflaton-driven
    evolution of the potential, vacuum ordering, and bubble
    dynamics of the spectator scalar field $s$.
    Initially, the $s=0$ vacuum is energetically favored.
    Near $N\simeq N_\star$, the nonzero field minimum
    becomes the lower energy state, allowing
    CDL tunneling~\cite{Coleman:1980aw,Sasaki:1994yj}
    and the nucleation
    of bubbles that subsequently expand during inflation.
    As the inflaton evolves away from the nucleation epoch,
    the vacuum ordering eventually reverses, rendering
    the bubble interiors energetically disfavored.
    The resulting reversal of the vacuum pressure
    opposes the outward motion of the bubble walls and,
    together with wall tension and Hubble damping,
    can gradually decelerate their
    expansion~\cite{Avelino:2008qy,Deng:2020mds}.
    By the end of inflation, the bubble interiors
    are assumed to remain in a positive energy
    metastable false vacuum state.
    (b) Post-inflationary evolution of a supercritical
    vacuum bubble. In the supercritical regime, an
    inflating interior develops behind a wormhole
    throat connecting it to the parent spacetime.
    The eventual closure of the throat leaves a PBH
    in the exterior
    region~\cite{Garriga:2015fdk,Deng:2017uwc}.
    (c) GWs induced by Poisson fluctuations
    in the PBH number density, provided that an early
    PBH dominated epoch occurs~\cite{Domenech:2020ssp}.
    (d) Baryogenesis during the final explosive stage of PBH
    evaporation. Energy deposition from Hawking radiation
    drives an expanding shock and creates a locally
    electroweak symmetric region. The enlarged view near the
    outward-propagating shock front illustrates the Higgs
    vacuum expectation value profile and the roles of
    CP-violating sources, chiral chemical potentials,
    and weak sphaleron processes in baryon number
    generation~\cite{Vanvlasselaer:2026vkh,Klipfel:2026nzx}.
    (e) GWs originating
    from direct graviton emission through Hawking
    radiation~\cite{Ireland:2023avg}.
    The GW spectra in panels (c) and (e) are shown
    schematically, with the indicated present-day peak
    frequencies corresponding to the characteristic
    values of the benchmark points considered in Sec.~\ref{sec:numerical_results}.
    This schematic figure was prepared with the assistance
    of generative AI tools and subsequently reviewed and
    refined by the authors.
    \\
    \par
    \endgroup
}

In this work, we investigate light PBH formation,
baryogenesis, and GW signatures within a common
cosmological framework based on a spectator sector FOPT
during inflation. Fig.~\ref{fig:PBH_mechanism}
provides a schematic overview of the proposed scenario.
We determine the vacuum tunneling history and the resulting
bubble population, follow the bubble growth and subsequent
supercritical evolution, and derive the corresponding PBH mass
and abundance from the spectator sector parameters and the
phenomenological bubble growth prescription. We then study
the Hawking evaporation of the resulting PBH population,
including its effects on the radiation and entropy evolution,
and calculate the baryon asymmetry generated during the final
explosive stage. We further investigate the GW signals from
direct graviton emission via Hawking radiation
and from PBH density fluctuations.

For the representative parameter points considered in
this work, the resulting PBHs have characteristic masses in
the range $\mathcal{O}(10^{5})$--$\mathcal{O}(10^{6})\,{\rm g}$.
For the benchmark CP-violating scale $\Lambda=1\,{\rm TeV}$,
the normalized baryon yield can reach
$\mathcal{O}(10^{-1})$ of the observed baryon asymmetry.
The GW signal induced by PBH density fluctuations
can reach a peak amplitude of
$\Omega_{\rm GW,0}^{\rm DF,peak}h^2\sim10^{-8}$,
where the superscript ``DF'' denotes the contribution
induced by density fluctuations.
The corresponding peak frequencies are of order
$10$--$10^2\,{\rm Hz}$,
whereas the GW contribution from direct graviton emission
peaks at much higher frequencies. These results illustrate how a single
spectator sector FOPT during inflation can
connect PBH formation with baryogenesis and GW signatures.

The remainder of this paper is organized as follows.
In Sec.~\ref{sec:spectator}, we introduce the spectator
sector model and study the FOPT and vacuum tunneling
during inflation.
In Sec.~\ref{sec:pbh_formation}, we discuss the evolution
of the nucleated bubbles and derive the resulting PBH
mass and abundance.
In Sec.~\ref{sec:evaporation}, we describe Hawking
evaporation and the associated cosmological evolution.
In Sec.~\ref{sec:baryogenesis}, we study baryogenesis
from the final explosive stage of PBH evaporation.
The associated GW signatures are discussed in
Sec.~\ref{sec:gravitational_waves}, and the numerical
results are presented in Sec.~\ref{sec:numerical_results}.
We conclude in Sec.~\ref{sec:conclusions}.
In Appendix~\ref{app:vacuum_decay}, we provide further details on vacuum
decay and stability during and after inflation.
Appendix~\ref{app:benchmark_parameters} lists the spectator potential parameters
for the benchmark points (BPs) considered in our numerical analysis. Throughout this work, we use natural units
$\hbar=c=k_B=1$ and denote the reduced Planck mass by
$M_{\rm Pl}\equiv(8\pi G)^{-1/2}$,
where $G$ is Newton's gravitational constant.

\section{Spectator sector phase transition during inflation}
\label{sec:spectator}

In this section, we introduce the spectator sector model
and investigate the inflaton-driven evolution of its vacuum
structure, the resulting bubble nucleation during inflation,
and the stability of the metastable bubble interiors.
The corresponding evolution of the spectator potential
and bubble dynamics is illustrated schematically in
Fig.~\ref{fig:PBH_mechanism}(a).

\subsection{Spectator sector model and vacuum structure}
\label{subsec:spectator_model}

We consider an inflationary scenario involving the inflaton $\phi$
and a spectator scalar field $s$, with the total potential given by
\begin{equation}
    V(\phi,s)
    =
    V_{\rm inf}(\phi)+U(s;\phi),
\end{equation}
where $V_{\rm inf}(\phi)$ is the inflaton potential
and $U(s;\phi)$ denotes the inflaton-dependent
spectator potential.
The spectator contribution is assumed to remain subdominant throughout the
relevant stage of inflation,
\begin{equation}
    |U(s;\phi)|\ll V_{\rm inf}(\phi),
\end{equation}
so that the background expansion is controlled by the inflaton sector. We take the spectator potential to be
\begin{equation}
    U(s;\phi)
    =
    \frac{1}{2}m_{\rm eff}^2(\phi)s^2
    -\frac{b}{3}s^3
    +\frac{\kappa}{4}s^4,
    \qquad
    b>0,
    \qquad
    \kappa>0.
    \label{eq:spectator_potential}
\end{equation}
Here $m_{\rm eff}^2(\phi)$ is the inflaton-dependent
spectator mass parameter, $b$ is a dimensionful cubic
coupling, and $\kappa$ is a dimensionless quartic coupling. Field-dependent masses can arise when scalar fields
are coupled to a rolling background field.
In particular, when a scalar field evolves through
special regions in field space, such as enhanced symmetry points, additional degrees of freedom can
become light and be produced nonadiabatically~\cite{Kofman:2004yc,Enomoto:2013mla,Kikuchi:2023uqo}. The backreaction of the
produced particles can significantly affect the dynamics of the inflaton and
provide an additional source of friction during inflation~\cite{Kofman:2004yc,Green:2009ds,Pearce:2016qtn}.

Motivated by these mechanisms, we consider an inflaton-dependent mass for the
spectator field whose variation becomes significant over a finite interval of
the inflationary trajectory. In contrast to trapped-inflation scenarios, we do
not rely on particle production backreaction to control the inflationary
dynamics. Although various inflationary models have been proposed
(see, e.g., Ref.~\cite{Kouniatalis:2025orn}),
we do not specify the inflaton potential explicitly in the
present setup. Instead, we assume that the inflaton continues
to evolve through the relevant field range without becoming trapped.

Any spectator particles produced by the nonadiabatic evolution of the mass
parameter are assumed to decay into lighter dark sector states. As discussed
in Ref.~\cite{Kofman:2004yc}, decay or annihilation of the produced particles
reduces their number density and can thereby weaken the trapping effect
associated with their backreaction. We assume that the decay dynamics lies in
a regime in which the spectator particle population is sufficiently depleted
for the inflaton to continue evolving through the relevant field range, while
the total energy density carried by the produced spectator particles and their
dark sector decay products remains subdominant,
\begin{equation}
    \rho_{\delta s}+\rho_D
    \ll
    3M_{\rm Pl}^2H_i^2.
\end{equation}
Here $\rho_{\delta s}$ and $\rho_D$ denote the energy
densities of the produced spectator particles and their
dark sector decay products, respectively, while $H_i$
is the approximately constant Hubble rate during
the relevant stage of inflation. The precise condition for satisfying these requirements depends on the
particle production and decay processes as well as on the inflaton potential
and the interactions between the inflaton and spectator sectors~\cite{Kofman:2004yc,Green:2009ds,Pearce:2016qtn}. Since we do not specify the
inflaton potential or a complete microscopic realization of these
interactions, a quantitative analysis of this condition is beyond the scope
of the present work and is left for future study.

 We adopt the
following phenomenological parametrization:
\begin{equation}
    m_{\rm eff}^2(\phi)
    =
    m_0^2
    -
    \Delta m^2
    \exp\left[
        -\frac{(\phi-\phi_\star)^2}{2\sigma_\phi^2}
    \right].
    \label{eq:mass_profile_phi}
\end{equation}
Here $m_0^2$ denotes the baseline value of the spectator
mass parameter, $\Delta m^2$ controls the amplitude of
its variation, and $\phi_\star$ and $\sigma_\phi$
specify the center and width of the Gaussian profile
in field space, respectively.

We parameterize the inflationary trajectory by the number of $e$-folds before
the end of inflation,
\begin{equation}
    N(t)
    =
    \int_t^{t_{\rm end}}H(t')\,dt',
    \qquad
    \frac{dN}{dt}=-H.
\end{equation}
Over the relevant inflationary interval, we assume a slowly varying Hubble rate, such that $H(N)\simeq H_\star\simeq H_i$.
Following Ref.~\cite{Cheng:2026npt}, we parameterize the inflaton evolution
during the relevant stage of inflation in terms of the number of $e$-folds,
with $\phi(N)\simeq NM_\mathrm{Pl}\sqrt{2\bar{\epsilon}_V}$ under the slow-roll
approximation. We then describe the corresponding evolution of the spectator
mass parameter as
\begin{equation}
m^2(N)
=
m_0^2
-
\Delta m^2
\exp\left[
-\frac{(N-N_\star)^2}{2\sigma_N^2}
\right],
\label{eq:mass_profile_N}
\end{equation}
where $N_\star$ specifies the center of the Gaussian profile and $\sigma_N$
characterizes its width in $e$-fold space. For convenience, we use the shorthand notation
$m^2(N)\equiv m_{\rm eff}^2[\phi(N)]$ and
$U_N(s)\equiv U[s;\phi(N)]$ throughout the following
discussion. Using the slow-roll relation between
the inflaton field and the number of $e$-folds, the corresponding width in
field space is
\begin{equation}
    \sigma_\phi
    \simeq
    \sqrt{2\bar{\epsilon}_V}\,M_{\rm Pl}\sigma_N.
\end{equation}
The evolution of $m^2(N)$ changes the relative depths of the two minima of the
spectator potential. To characterize the resulting vacuum structure, we first
note that the two minima become degenerate at
\begin{equation}
    m_{\rm deg}^2
    =
    \frac{2b^2}{9\kappa}.
    \label{eq:mdeg}
\end{equation}
It is convenient to introduce the dimensionless parameters
\begin{equation}
    \alpha_0
    \equiv
    \frac{m_0^2}{m_{\rm deg}^2},
    \qquad
    \delta_m
    \equiv
    \frac{\Delta m^2}{m_{\rm deg}^2},
\end{equation}
together with
\begin{equation}
    \delta_N
    \equiv
    1-\frac{m^2(N)}{m_{\rm deg}^2}
    =
    1-\alpha_0
    +
    \delta_m
    \exp\left[
        -\frac{(N-N_\star)^2}{2\sigma_N^2}
    \right].
    \label{eq:delta_N}
\end{equation}
At fixed $N$, the stationary points of the spectator potential satisfy
\begin{equation}
    \frac{\partial U_N}{\partial s}
    =
    s\left[m^2(N)-bs+\kappa s^2\right]
    =
    0.
\end{equation}
Besides the stationary point at $s=0$, the two nonzero stationary points are
\begin{equation}
    s_{\pm,N}
    =
    \frac{b}{6\kappa}
    \left(
        3\pm\sqrt{1+8\delta_N}
    \right),
    \label{eq:spectator_extrema}
\end{equation}
where $s_{+,N}$ is the nonzero local minimum and $s_{-,N}$ is the
barrier top within the parameter region of interest.
The two nonzero extrema exist for $\delta_N>-1/8$.
The relative vacuum energies are controlled by $\delta_N$.
For $\delta_N>0$, the nonzero minimum has a lower vacuum energy
than the $s=0$ configuration. In the range $0<\delta_N<1$,
the $s=0$ configuration remains a local minimum and can
therefore decay to the nonzero phase through bubble nucleation.
In contrast, for $-1/8<\delta_N<0$, the $s=0$ phase is the
true vacuum, while the nonzero minimum remains as a
positive energy metastable state.

At the center of the Gaussian profile, $N=N_\star$, we define
$\delta_\star\equiv\delta_{N_\star}=1-\alpha_0+\delta_m$.
For the desired vacuum transition, we require
$0<\delta_\star<1$, such that the vacuum at $s=0$ remains
a local minimum while the nonzero minimum becomes
energetically favored. Tunneling from the vacuum at $s=0$
to the nonzero minimum can therefore occur around $N=N_\star$.

By the end of inflation, $N=0$, the mass parameter has evolved back toward
its baseline value. We define $\delta_{\rm end}\equiv\delta_{N=0}$
and require $-1/8<\delta_{\rm end}<0$, such that the vacuum
at $s=0$ is again the global minimum, while the nonzero local minimum remains
as a positive energy metastable state. We assume that the field inside the
bubbles remains near this nonzero local minimum after the ordering of the two
vacua is reversed. The bubbles nucleated during the earlier interval can
therefore retain a metastable interior at the end of inflation.

Throughout this work, the spectator sector FOPT refers to this finite interval
of vacuum tunneling and bubble nucleation during inflation. We do not require the phase transition to complete through the percolation of nucleated bubbles.
 Instead, we consider a
dilute bubble population for which bubble overlap is negligible and whose
subsequent evolution can lead to PBH formation.

\subsection{Vacuum tunneling during inflation}
\label{subsec:inflationary_tunneling}

We next determine the vacuum decay rate during the interval in which the
nonzero phase is energetically favored. Since the spectator potential varies
with the inflationary background, the tunneling problem is intrinsically
time dependent. We adopt a quasi-static approximation in which the background
value of $\phi$, or equivalently $N$, is held fixed during each tunneling
event. 

At each fixed $N$, we evaluate the CDL tunneling exponent
for the spectator potential in
Eq.~\eqref{eq:spectator_potential}, with the mass parameter
specified by Eq.~\eqref{eq:mass_profile_N}, using the
semi-analytic treatment of quartic potentials in de Sitter
space developed by Sasaki, Stewart, and Tanaka
(SST)~\cite{Sasaki:1994yj}.
For our potential, the tunneling exponent can be written as
\begin{equation}
    B_{\rm CDL}(N)
    =
    \frac{\pi^2}{6\kappa}
    (1-\delta_N)
    \widetilde B_{\rm SST}
    \left[
        \varepsilon_{\rm SST}(N),
        \Omega_{\rm SST}(N)
    \right],
    \label{eq:B_CDL}
\end{equation}
where the dimensionless variables and the fitting function
$\widetilde B_{\rm SST}$ are summarized in
Appendix~\ref{app:vacuum_decay}.

We focus on the parameter region in which the transition is described by a
nonconstant CDL bounce rather than by Hawking--Moss transition~\cite{Weinberg:2006pc,Hawking:1981fz}. The curvature
at the barrier top is
\begin{equation}
    \mu_{\rm top}^2(N)
    \equiv
    -U_N''(s_{-,N})
    =
    \frac{b^2}{18\kappa}
    q_N(3-q_N),
    \qquad
    q_N\equiv\sqrt{1+8\delta_N},
\end{equation}
and at the center of the nucleation feature we impose~\cite{Cheng:2026npt,Joti:2017fwe}
\begin{equation}
    \frac{\mu_{{\rm top},\star}^2}{H_i^2}
    \gtrsim 4.
    \label{eq:HM_veto}
\end{equation}
Together with the weak gravitational backreaction condition required for the
quartic fit, Eq.~\eqref{eq:HM_veto} defines the region in which we use the CDL
description.

The nucleation rate per unit physical four-volume is then written as~\cite{Coleman:1977py,Callan:1977pt,Coleman:1980aw}
\begin{equation}
    \Gamma(N)
    =
    A_{\rm tun}(N)
    \exp[-B_{\rm CDL}(N)],
    \label{eq:nucleation_rate}
\end{equation}
where $A_{\rm tun}$ denotes the fluctuation prefactor. For the numerical
estimates below, we use $A_{\rm tun}\simeq H_i^4$ as an order-of-magnitude
normalization and retain the exponential dependence on the bounce action as
the dominant source of the nucleation profile.

\subsection{Bubble nucleation rate and source distribution}
\label{subsec:bubble_source}

The time dependence of the spectator mass localizes vacuum tunneling around
$N=N_\star$. In the parameter region of interest, the CDL exponent reaches
its minimum near this point, and the corresponding nucleation rate is therefore
sharply peaked within a finite interval.

To determine the resulting bubble abundance, we normalize the scale factor
such that $a(t_{\rm end})=1$, which gives $a(N)=e^{-N}$ during inflation.
For a unit comoving volume, the corresponding physical volume at the time of
nucleation is $a^3(N)$. The number of bubbles nucleated during an interval
$dN$ is therefore
\begin{equation}
    dn_{\rm bub}^{\rm co}
    =
    a^3(N)\Gamma(N)|dt|
    =
    a^3(N)\frac{\Gamma(N)}{H(N)}\,dN
    =
    \frac{e^{-3N}\Gamma(N)}{H(N)}\,dN,
\end{equation}
where we have used $dN/dt=-H$. It follows that
\begin{equation}
    \frac{dn_{\rm bub}^{\rm co}}{dN}
    =
    \frac{e^{-3N}\Gamma(N)}{H(N)}.
    \label{eq:bubble_source_distribution}
\end{equation}
The total comoving number density of nucleated bubbles is therefore
\begin{equation}
    n_{\rm bub}^{\rm co}
    =
    \int_{\rm source} dN\,
    \frac{e^{-3N}\Gamma(N)}{H(N)}.
    \label{eq:bubble_number_density}
\end{equation}
For later convenience, we characterize the localized source by its peak
quantities,
\begin{equation}
    \Gamma_\star
    \equiv
    \Gamma(N_\star),
    \qquad
    H_\star
    \equiv
    H(N_\star),
\end{equation}
and define an effective source width,
\begin{equation}
   \Delta N_{\rm nuc}
    \equiv
    \int_{\rm source} dN\,
    e^{-3(N-N_\star)}
    \frac{H_\star}{H(N)}
    \frac{\Gamma(N)}{\Gamma_\star}.
    \label{eq:DeltaN_eff}
\end{equation}
The total comoving bubble number density can then be written as
\begin{equation}
    n_{\rm bub}^{\rm co}
    =
    e^{-3N_\star}
    \frac{\Gamma_\star}{H_\star}
    \Delta N_{\rm nuc}.
    \label{eq:bubble_source_peak_width}
\end{equation}
Here, $\Delta N_{\rm nuc}$ denotes the effective width of the nucleation
interval, with the $N$ dependence of the factor $e^{-3N}$ and of the Hubble
rate included in its definition.

Since the nucleation rate is localized around $N_\star$, the resulting bubble
population is also concentrated around a characteristic comoving scale. An
approximately monochromatic PBH population is obtained when the corresponding
variation of the PBH mass across the source interval is sufficiently small.
We quantify this condition after deriving the relation between the nucleation
time and the PBH mass in Sec.~\ref{sec:pbh_formation}.

\subsection{Vacuum structure at the end of inflation and stability of the bubble interior}
\label{subsec:bubble_stability}

As the spectator mass returns toward its baseline value near the end of
inflation, the vacuum at $s=0$ becomes the global minimum again, while the
nonzero local minimum remains present for $-1/8<\delta_{\rm end}<0$, as
discussed in Sec.~\ref{subsec:spectator_model}. For the subsequent evolution,
we focus on the regime in which the field configuration inside a nucleated
bubble remains sufficiently close to the nonzero local minimum,
$s_{\rm in,end}\simeq s_{+,{\rm end}}$. The bubble interior therefore
corresponds to a metastable vacuum with positive energy density. The associated
vacuum energy density is
\begin{equation}
    \rho_b
    =
    \frac{b^4}{5184\kappa^3}
    (1-q_{\rm end})(3+q_{\rm end})^3,
    \qquad
    q_{\rm end}
    =
    \sqrt{1+8\delta_{\rm end}},
    \label{eq:rho_b}
\end{equation}
which defines the corresponding interior de Sitter scale,
\begin{equation}
    H_b^2
    =
    \frac{\rho_b}{3M_{\rm Pl}^2}.
    \label{eq:H_b}
\end{equation}

For the bubbles to survive and subsequently undergo the
supercritical evolution discussed in Sec.~\ref{sec:pbh_formation},
reverse vacuum tunneling from their metastable interiors
to the true vacuum must remain sufficiently suppressed.  During the remaining inflationary
evolution, we estimate its decay in the quasi-de Sitter background using the
quartic vacuum decay treatment of Ref.~\cite{Sasaki:1994yj}. After reheating,
we assume that the spectator sector remains decoupled from the ambient
radiation bath, so that thermal corrections to the spectator potential are
negligible. The decay of the metastable interior can then be treated as a
zero temperature $O(4)$ tunneling process, for which we use the quartic potential
results of Ref.~\cite{Adams:1993zs}. The corresponding bounce scales and the
validity of the flat space approximation are discussed in
Appendix~\ref{app:vacuum_decay}.

Rather than attempting a fully time-dependent calculation of the integrated
decay probability, we impose conservative benchmark conditions requiring the
vacuum decay evaluated at the end of inflation, $N=0$, to be exponentially
more suppressed than the tunneling process responsible for bubble production,
\begin{equation}
    B_{\rm end}^{\rm dS}
    >
    B_\star+\Delta B_{\rm safe},
    \qquad
    B_{4,\rm end}^{\rm flat}
    >
    B_\star+\Delta B_{\rm safe},
    \qquad
    \Delta B_{\rm safe}\gg1,
    \label{eq:endpoint_stability}
\end{equation}
whenever the corresponding decay channel is physically applicable. Here $B_\star\equiv B_{\rm CDL}(N_\star)$ is the
characteristic tunneling exponent for bubble nucleation,
while $B_{\rm end}^{\rm dS}$ and $B_{4,\rm end}^{\rm flat}$
denote the reverse tunneling exponents evaluated in the
de Sitter and flat-space approximations, respectively.
The parameter $\Delta B_{\rm safe}$ specifies the required
additional exponential suppression of reverse vacuum tunneling.
In our numerical analysis, we adopt the benchmark value
$\Delta B_{\rm safe}=10$ throughout the parameter scan. These
conditions are used as benchmark stability criteria and are not intended to
replace a complete treatment of vacuum decay throughout the subsequent
time-dependent background.

Together with the bubble source derived in
Sec.~\ref{subsec:bubble_source}, the interior de Sitter scale
$H_b$ defined in Eq.~\eqref{eq:H_b} provides the initial
conditions relevant for the subsequent bubble evolution.

\section{PBH formation from bubbles}
\label{sec:pbh_formation}

The preceding section determines both the bubble source and the metastable
vacuum energy stored in the bubble interior. We now follow the
subsequent evolution of these bubbles, determine the condition for entering the
supercritical regime, and map the inflationary nucleation history onto the PBH
mass and abundance. For simplicity, we assume instantaneous reheating followed by
radiation domination. The bubble wall evolution and the effects
of the surrounding plasma are treated phenomenologically.

\subsection{Inflationary evolution and bubble size}
\label{subsec:bubble_evolution}

Consider a bubble nucleated at $N$ $e$-folds before the end of inflation with
physical radius $R_n(N)$. Since the scale factor is normalized by
$a(t_{\rm end})=1$, the scale factor at nucleation is $a_n=e^{-N}$. The
comoving bubble radius at the end of inflation can therefore be written as
\begin{equation}
    \chi_i(N)
    =
    e^N R_n(N)
    +
    \int_{t_n}^{t_{\rm end}}
    \frac{v_w(t')}{a(t')}\,dt',
    \label{eq:bubble_radius_general}
\end{equation}
where $t_n$ is the bubble nucleation time and $v_w$
denotes the wall velocity relative to the Hubble flow.
For an approximately constant inflationary Hubble rate $H_i$, we parameterize
the integrated wall propagation by an effective velocity $v_{\rm eff}$ through
\begin{equation}
    \int_{t_n}^{t_{\rm end}}
    \frac{v_w(t')}{a(t')}\,dt'
    \equiv
    \frac{v_{\rm eff}}{H_i}
    \left(e^N-1\right).
    \label{eq:v_eff_definition}
\end{equation}
The resulting scale is thus
\begin{equation}
    \chi_i(N)
    =
    e^N R_n(N)
    +
    \frac{v_{\rm eff}}{H_i}
    \left(e^N-1\right).
    \label{eq:bubble_radius_end}
\end{equation}
The bubble size at the end of inflation is therefore determined
by both the cosmological expansion of its initial radius and
the propagation of the bubble wall during inflation.
Similar inflationary stretching of nucleated vacuum bubbles and
their later evolution into compact objects have been studied in
Ref.~\cite{Garriga:2015fdk}.

In this work, we consider the regime in which the second term in
Eq.~\eqref{eq:bubble_radius_end} dominates over the stretched nucleation radius,
\begin{equation}
    e^N R_n(N)
    \ll
    \frac{v_{\rm eff}}{H_i}
    \left(e^N-1\right),
    \label{eq:Rn_negligible_condition}
\end{equation}
so that
\begin{equation}
    \chi_i(N)
    \simeq
    \frac{v_{\rm eff}}{H_i}
    \left(e^N-1\right).
    \label{eq:bubble_radius_approx}
\end{equation}
For the characteristic nucleation time $N=N_\star$, we denote
$\chi_i\equiv\chi_i(N_\star)$. Since $a(t_{\rm end})=1$, this comoving scale is
numerically equal to the physical bubble radius at the end of inflation. We further adopt, as a baseline approximation, that the characteristic
comoving size of the bubble is determined predominantly during inflation.

The evolution of the bubble wall is expected to proceed through qualitatively
different stages as the relative vacuum energies vary during inflation. Around
the nucleation epoch, the nonzero minimum lies below the vacuum at $s=0$, and
the corresponding vacuum energy difference provides an outward volume pressure
that drives the bubble wall away from the nucleation center. The wall can
therefore acquire a sizable velocity during this stage, although its
motion is simultaneously affected by the wall tension and curvature as well as
by Hubble damping. As the spectator potential subsequently evolves and the
vacuum ordering is reversed, the nonzero phase inside the bubble becomes the
higher energy metastable vacuum. The vacuum pressure contribution then reverses
sign and acts against the outward motion of the wall. Together with the
curvature force and cosmological damping, this generically leads to a
deceleration of the outward  motion, as is familiar from the dynamics
of biased domain walls and inflationary vacuum bubbles~\cite{Avelino:2008qy,Deng:2020mds}.

In the parameter region considered here, however, the post-reversal vacuum
bias cannot be arbitrarily large. In particular, the stability conditions
imposed on the metastable bubble interior require its decay exponent at the end
of inflation to be substantially larger than the exponent associated with the
earlier bubble nucleation epoch. Within the present spectator potential, this
requirement disfavors configurations with an excessively strong
post-reversal vacuum energy splitting. Consequently, although the reversal of
the vacuum ordering provides a robust mechanism for decelerating an
outward-moving wall, the deceleration need not be instantaneous and can occur
over a finite interval of the inflationary evolution. The actual trajectory
depends on the interplay of the evolving vacuum structure, the cosmological
background, and the bubble wall dynamics. A quantitative determination
therefore requires solving the full time-dependent wall equation of motion.

In the present phenomenological treatment, we do not attempt to resolve this
detailed evolution. Instead, the accumulated wall propagation during inflation
is encoded in the effective velocity $v_{\rm eff}$ introduced above. Motivated
by the expected deceleration following the reversal of the vacuum ordering, we
focus on the phenomenological regime in which the residual velocity is
small by the end of inflation,
\begin{equation}
    v_w(t_{\rm end})\simeq0.
\end{equation}
This should be regarded as a phenomenologically motivated regime rather than as
a first-principles prediction of the present analysis. Importantly, a small
terminal velocity does not imply a small $v_{\rm eff}$: the latter
measures the integrated displacement of the wall over the entire inflationary
interval and can therefore remain sizable if the bubble undergoes substantial
outward propagation during the earlier stage. The change in the sign of the
vacuum pressure, together with the additional damping effects discussed above,
provides physical support for such a separation between the accumulated wall
propagation and its residual velocity at the end of inflation.

Following the treatment of inflationary vacuum bubbles in
Ref.~\cite{Garriga:2015fdk}, we use $\chi_i$ as the reference
comoving scale for determining the horizon re-entry time.
This prescription is consistent with the phenomenological
regime adopted here, in which the residual motion of the
wall is already small at the end of inflation.
A fully self-consistent treatment of the time-dependent
wall dynamics, including the evolution of the spectator
potential, gravitational effects, and the transition through
reheating, is left for future work.

\subsection{Horizon re-entry and supercritical bubble evolution}
\label{subsec:supercritical_bubbles}

After instantaneous reheating, we assume radiation domination with the Hubble
rate matched to $H_i$ at $a=1$. The subsequent expansion is then described by
\begin{equation}
    H(a)
    =
    H_i a^{-2}.
    \label{eq:H_radiation}
\end{equation}
A bubble scale $\chi_i(N)$ re-enters the Hubble horizon when its physical size
$a_H\chi_i(N)$ becomes equal to $H^{-1}(a_H)$. Using
Eq.~\eqref{eq:H_radiation}, this gives
\begin{equation}
    a_H(N)
    =
    H_i\chi_i(N).
    \label{eq:horizon_entry_scale_factor}
\end{equation}
For the re-entry to occur after reheating, the relevant source must satisfy
$a_H\gtrsim1$. The Hubble parameter at re-entry is consequently
\begin{equation}
    H_{\rm form}(N)
    =
    \frac{H_i}{a_H^2(N)}
    =
    \frac{1}{H_i\chi_i^2(N)},
    \label{eq:H_form_N}
\end{equation}
and the corresponding Hubble time is
\begin{equation}
    t_H(N)
    \equiv
    H_{\rm form}^{-1}(N)
    =
    H_i\chi_i^2(N).
    \label{eq:tH_N}
\end{equation}
We use $t_H$ as the
characteristic horizon re-entry time scale throughout.

Whether a re-entering vacuum bubble lies on the supercritical branch is not
determined by the interior vacuum energy alone. In addition to the de Sitter
scale $H_b$ defined in Eq.~\eqref{eq:H_b}, the wall tension introduces the
gravitational scale~\cite{Garriga:2015fdk,Deng:2017uwc,Deng:2020mds}
\begin{equation}
    H_\Sigma
    \equiv
    2\pi G\Sigma,
    \label{eq:H_sigma}
\end{equation}
where $\Sigma$ is the wall tension. Following
Ref.~\cite{Garriga:2015fdk}, the critical mass separating the subcritical and
supercritical bubble regimes is parametrically determined by the interior
de Sitter scale and the gravitational scale associated with the wall tension,
\begin{equation}
    G M_{\rm cr}
    \sim
    \min\left\{
        H_b^{-1},
        H_\Sigma^{-1}
    \right\},
    \label{eq:critical_mass_supercritical}
\end{equation}
up to factors of order unity~\cite{Garriga:2015fdk}. Above this critical scale, the bubble can enter
the supercritical regime, in which the false vacuum
interior continues to inflate and a black hole
connected to the inflating interior through a
wormhole is seen from the parent universe~\cite{Garriga:2015fdk}.
This evolution is illustrated schematically
in Fig.~\ref{fig:PBH_mechanism}(b).

During radiation domination, the horizon mass associated
with the re-entry of a bubble nucleated at $N$ is
\begin{equation}
    M_H(N)
    =
    \frac{4\pi}{3}\rho_{\rm rad}(N)H_{\rm form}^{-3}(N)
    =
    \frac{1}{2G H_{\rm form}(N)}
    =
    \frac{t_H(N)}{2G}.
    \label{eq:horizon_mass_N}
\end{equation}

Using the horizon-mass estimate in
Eq.~\eqref{eq:horizon_mass_N}, the corresponding supercriticality condition
can be written parametrically as~\cite{Garriga:2015fdk,Deng:2017uwc,Deng:2020mds}
\begin{equation}
    t_H(N)
    \gtrsim
    \min\left\{
        H_b^{-1},
        H_\Sigma^{-1}
    \right\}.
    \label{eq:supercritical_condition}
\end{equation}

Since the criterion involving $H_\Sigma$ is based on the thin-wall
description, while some bubbles in our parameter space may lie outside the
regime where the thin wall approximation is reliable, we do not use
$H_\Sigma$ as a universal benchmark. Instead, we adopt the phenomenological benchmark condition
\begin{equation}
    t_H(N)\gtrsim H_b^{-1},
    \label{eq:supercritical_Hb_benchmark}
\end{equation}
which is sufficient to satisfy Eq.~\eqref{eq:supercritical_condition} since
$\min\{H_b^{-1},H_\Sigma^{-1}\}\leq H_b^{-1}$, but should not be interpreted
as an exact necessary and sufficient criterion.
\subsection{PBH mass and abundance}
\label{subsec:pbh_mass_abundance}

We now restrict our attention to bubbles satisfying the supercriticality
condition discussed above. Following Ref.~\cite{Garriga:2015fdk}, we estimate
the mass of the resulting PBH from the horizon mass associated with the
corresponding re-entry scale. Using Eq.~\eqref{eq:horizon_mass_N}, this gives,
up to factors of order unity,
\begin{equation}
    M_{\rm PBH}(N)
    \simeq
    M_H(N)
    =
    \frac{t_H(N)}{2G}
    =
    \frac{H_i\chi_i^2(N)}{2G}.
    \label{eq:PBH_mass_general_N}
\end{equation}
In terms of the commonly used parametrization
$M_{\rm PBH}=\gamma M_H$, where $\gamma$ accounts for the order-one relation
between the PBH mass and the horizon mass at formation, this prescription
corresponds to setting $\gamma=1$~\cite{Deng:2020mds,Deng:2017uwc,Garriga:2015fdk}. We therefore do not introduce an additional
collapse efficiency parameter. The supercriticality condition determines
whether a bubble follows the supercritical branch, while the characteristic
mass of the resulting PBH is estimated from the horizon scale associated with
the bubble size acquired during inflation.

When the initial nucleation radius can be neglected, inserting
Eq.~\eqref{eq:bubble_radius_approx} into
Eq.~\eqref{eq:PBH_mass_general_N} gives
\begin{equation}
    M_{\rm PBH}(N)
    \simeq
    \frac{v_{\rm eff}^2}{2G H_i}
    \left(e^N-1\right)^2.
    \label{eq:PBH_mass_N}
\end{equation}
Since the nucleation rate is concentrated around $N_\star$, the corresponding
characteristic PBH mass is
\begin{equation}
    M_\star
    \equiv
    M_{\rm PBH}(N_\star)
    \simeq
    M_H(N_\star)
    =
    \frac{H_i\chi_i^2}{2G}
    \simeq
    \frac{v_{\rm eff}^2}{2G H_i}
    \left(e^{N_\star}-1\right)^2.
    \label{eq:PBH_mass_star}
\end{equation}
The associated Hubble time at horizon re-entry is
\begin{equation}
    t_{H,\star}
    =
    H_{\rm form,\star}^{-1}
    =
    H_i\chi_i^2
    \simeq
    2G M_\star.
    \label{eq:PBH_horizon_time_star}
\end{equation}
Thus, within the approximations adopted here, both the characteristic PBH mass
and the corresponding formation time scale are determined by the same bubble
scale acquired during inflation.

The PBH abundance is inherited from the bubble nucleation history derived in
Sec.~\ref{subsec:bubble_source}. In the parameter region retained in our
analysis, the bubbles contributing appreciably to the nucleation abundance
satisfy the supercriticality condition, and we associate one exterior PBH with
each such bubble. The resulting comoving PBH number density is therefore
\begin{equation}
    n_{\rm PBH,\star}^{\rm co}
    \simeq
    n_{\rm bub}^{\rm co}
    =
    e^{-3N_\star}
    \frac{\Gamma_\star}{H_\star}
    \Delta N_{\rm nuc},
    \label{eq:PBH_comoving_number_density}
\end{equation}
where $\Delta N_{\rm nuc}$ is defined in
Eq.~\eqref{eq:DeltaN_eff}. At horizon entry,
\begin{equation}
    a_{H,\star}
    =
    H_i\chi_i,
\end{equation}
so that the physical PBH number density is
\begin{equation}
    n_{\rm PBH,\star}^{H}
    =
    \frac{n_{\rm PBH,\star}^{\rm co}}
         {a_{H,\star}^3}.
    \label{eq:PBH_physical_number_density}
\end{equation}
The corresponding PBH-to-radiation energy density ratio is
\begin{equation}
    \kappa_{H}^{\rm SB}
    \equiv
    \frac{\rho_{\rm PBH}^{H}}
         {\rho_{\rm rad}^{H}}
    \simeq
    \frac{
        n_{\rm PBH,\star}^{\rm co}M_\star
    }{
        a_{H,\star}^3
        \left(3M_{\rm Pl}^2H_{\rm form,\star}^2\right)
    }.
    \label{eq:PBH_fraction_formation}
\end{equation}
Here and throughout this work, the superscript or
subscript ``SB'' stands for ``spectator bubble''
and labels quantities associated with the PBH
population originating from the spectator sector
phase transition. The superscript ``co'' denotes comoving quantities,
while the superscript $H$ indicates physical quantities
evaluated at horizon re-entry.
A subscript $\star$ denotes quantities associated with
the characteristic nucleation epoch $N=N_\star$.
The ratio $\kappa_H^{\rm SB}$ is a derived quantity
rather than an independent normalization of the
PBH abundance,
which is fixed by the integrated tunneling history
through Eq.~\eqref{eq:PBH_comoving_number_density}.

Within the phenomenological framework adopted here,
the interior vacuum energy sets the characteristic
scale entering our benchmark supercriticality condition,
whereas the PBH mass is primarily governed by the horizon
scale associated with the bubble size acquired during
inflation~\cite{Garriga:2015fdk,Deng:2017uwc,Deng:2020mds}.
The PBH abundance, in turn, is determined by the
vacuum tunneling history.

\subsection{Monochromatic approximation}
\label{subsec:monochromatic_source}

We emphasize that the quantity $\Delta N_{\rm nuc}$ introduced above is an
effective width defined through the integrated nucleation abundance and should
not be identified with the physical width of the nucleation interval relevant
for the PBH mass distribution. We denote by
$\Delta N_{\rm PBH}$ the range in $N$ over which bubble nucleation contributes
appreciably to the PBH population,
\begin{equation}
N\in[N_-,N_+],
\qquad
\Delta N_{\rm PBH}
\equiv
N_+-N_-.
\label{eq:DeltaN_PBH_definition}
\end{equation}
Within the mass mapping of Eq.~\eqref{eq:PBH_mass_N}, the masses associated
with the two ends of this interval satisfy
\begin{equation}
\frac{M_{\rm max}}{M_{\rm min}}
=
\left[
\frac{e^{N_+}-1}
{e^{N_-}-1}
\right]^2.
\label{eq:PBH_mass_width_exact}
\end{equation}
For the parameter region relevant to our analysis, nucleation occurs at
$N\gg1$, such that $e^N\gg1$. Eq.~\eqref{eq:PBH_mass_width_exact}
therefore reduces to
\begin{equation}
\frac{M_{\rm max}}{M_{\rm min}}
\simeq
e^{2(N_+-N_-)}
=
e^{2\Delta N_{\rm PBH}},
\label{eq:PBH_mass_width_largeN}
\end{equation}
or equivalently,
\begin{equation}
\Delta\ln M_{\rm PBH}
\equiv
\ln\frac{M_{\rm max}}{M_{\rm min}}
\simeq
2\Delta N_{\rm PBH}.
\label{eq:PBH_log_mass_width}
\end{equation}

A localized PBH-producing nucleation history is compatible with the
phenomenological bubble wall evolution adopted in
Sec.~\ref{subsec:bubble_evolution}. In particular, the condition
$v_w(t_{\rm end})\simeq0$ constrains the residual velocity of the wall
at the end of inflation, whereas $v_{\rm eff}$ characterizes the accumulated
wall propagation over the preceding inflationary evolution. The spectator
potential contains sufficient parametric freedom to vary the vacuum energy
difference around the nucleation epoch and after the reversal of the vacuum
ordering. A bubble can therefore acquire a sizable outward displacement
during the earlier stage and subsequently decelerate over a finite interval
without requiring the PBH-producing nucleation history itself to extend over
the same interval. The corresponding freedom in the spectator potential is
discussed further in Appendix~\ref{app:benchmark_parameters}.

Within the phenomenological treatment adopted here, the post-reversal vacuum
structure affects the subsequent wall evolution and the selection of the
supercritical branch, but does not enter the PBH mass mapping directly. Once
the accumulated propagation is parameterized by $v_{\rm eff}$, the
characteristic mass in Eq.~\eqref{eq:PBH_mass_N} is determined by the
nucleation time, $H_i$, and $v_{\rm eff}$. Adjusting the vacuum energy
difference after nucleation in order to realize a small residual wall velocity
therefore does not by itself require a broad PBH mass distribution.

Moreover, the duration over which the spectator potential evolves need not
coincide with the width of the nucleation history. The bubble source is
controlled by
\begin{equation}
\frac{dn_{\rm bub}^{\rm co}}{dN}
=
\frac{e^{-3N}\Gamma(N)}{H(N)},
\end{equation}
with
\begin{equation}
\Gamma(N)
\propto
\exp[-B_{\rm CDL}(N)].
\end{equation}
Since the tunneling exponent develops a minimum around $N_\star$ in the
parameter region of interest, the exponential dependence of the nucleation
rate can localize the actual bubble source much more strongly than the
underlying variation of the spectator potential~\cite{Wang:2025hwc}. A finite interval of
potential evolution can therefore remain compatible with a substantially more
localized nucleation history relevant for PBH production. This also explains
why the width controlling the PBH mass distribution should be characterized
directly by $\Delta N_{\rm PBH}$ rather than by the width of the spectator
potential or by the effective abundance parameter $\Delta N_{\rm nuc}$.

Motivated by these considerations, we adopt
$\Delta N_{\rm PBH}\lesssim 1$ as a phenomenological
condition for the monochromatic limit treatment.
This condition is distinct from the benchmark choice
$\Delta N_{\rm nuc}=1$ adopted in
Sec.~\ref{sec:numerical_results}, where
$\Delta N_{\rm nuc}$ characterizes the effective
normalization of the integrated bubble abundance.
From Eq.~\eqref{eq:PBH_mass_width_largeN}, it implies
$M_{\rm max}/M_{\rm min}\lesssim e^2\simeq 7.4$,
or equivalently, $\Delta\ln M_{\rm PBH}\lesssim 2$.
The PBHs contributing appreciably to the subsequent evolution are therefore
confined to less than one decade in mass. In addition, within the present
approximation the inflationary Hubble rate is slowly varying, the accumulated
wall propagation is described by the common parameter $v_{\rm eff}$, and the
initial nucleation radius is neglected in the mass mapping. The dominant
source of the PBH mass variation is therefore the finite range of nucleation
times itself.

For the benchmark analysis below, we represent this localized PBH population
by its characteristic mass $M_\star$ and adopt the monochromatic
approximation,
\begin{equation}
\frac{dn_{\rm PBH,i}^{\rm co}}{dM_i}
\simeq
n_{\rm PBH,\star}^{\rm co}
\delta\left(M_i-M_\star\right).
\label{eq:PBH_monochromatic_source}
\end{equation}
Here $M_i$ denotes the initial PBH mass at formation. This prescription should be understood as the monochromatic-limit
representation of a finite but localized PBH mass distribution. Possible
effects associated with its residual finite width are discussed in
Sec.~\ref{sec:conclusions}.

\section{Hawking evaporation and cosmological evolution}
\label{sec:evaporation}

The preceding section determines the characteristic PBH mass $M_\star$, the
absolute comoving number density $n_{\rm PBH,\star}^{\rm co}$, and the scale
factor $a_{H,\star}$ at horizon re-entry. We now follow the subsequent Hawking
evaporation of this nearly monochromatic PBH population and its impact on the
cosmological background. Throughout this section, we retain the global
normalization $a(t_{\rm end})=1$ introduced in
Sec.~\ref{sec:pbh_formation}, without rescaling the scale factor at PBH
formation. We take the horizon re-entry of the characteristic bubble scale as
the initial epoch of the subsequent evolution and denote the elapsed time from
this epoch by $\Delta t$.

\subsection{Hawking evaporation}
\label{subsec:hawking_evaporation}

We consider uncharged and nonrotating PBHs. The Hawking temperature of a
Schwarzschild black hole of mass $M$ is~\cite{Hawking:1975vcx}
\begin{equation}
    T_H(M)
    =
    \frac{1}{8\pi G M}
    =
    \frac{M_{\rm Pl}^2}{M}.
    \label{eq:Hawking_temperature}
\end{equation}
As the PBH evaporates, its mass decreases and its Hawking temperature rises,
leading to an increasingly rapid energy loss. For the mass range relevant to
the baryogenesis mechanism considered here,
$M_{\rm PBH}\lesssim10^8\,{\rm g}$, one has
$T_H\gtrsim10^5\,{\rm GeV}$, so that all Standard Model particle species can
be efficiently emitted~\cite{Vanvlasselaer:2026vkh,Klipfel:2026nzx}. We
therefore adopt the high temperature approximation used in
Refs.~\cite{Vanvlasselaer:2026vkh,Klipfel:2026nzx}, for which the total mass-loss rate due to Hawking
evaporation is
\begin{equation}
    -\frac{dM}{dt}
    =
    \frac{\mathcal{A}_{\rm evap} f_{\max}}{M^2},
    \qquad
    \mathcal{A}_{\rm evap}
    =
    6.04\times10^{72}\,{\rm GeV}^4,
    \qquad
    f_{\max}=15.52.
    \label{eq:hawking_mass_loss}
\end{equation}
Here $f_{\max}$ is the limiting value of the Page factor~\cite{Page:1976df} when all Standard
Model degrees of freedom are kinematically accessible. Since the PBHs studied
in this work remain well within the regime in which this approximation applies,
we take $f(M)\simeq f_{\max}$ throughout the subsequent evaporation
calculation.

For later convenience, we define
$\mathcal{A}_f\equiv\mathcal{A}_{\rm evap}f_{\max}$. For the
monochromatic population in Eq.~\eqref{eq:PBH_monochromatic_source}, all PBHs
are formed with the same initial mass $M_\star$. Measuring the elapsed time
from horizon re-entry by $\Delta t$, with $\Delta t=0$ at this epoch,
integration of Eq.~\eqref{eq:hawking_mass_loss} gives
\begin{equation}
    M(\Delta t)
    =
    \left(
        M_\star^3-3\mathcal{A}_f\Delta t
    \right)^{1/3},
    \qquad
    0\leq\Delta t<\tau_\star,
    \label{eq:PBH_mass_evolution}
\end{equation}
where
\begin{equation}
    \tau_\star
    =
    \frac{M_\star^3}{3\mathcal{A}_f}
    \label{eq:PBH_lifetime}
\end{equation}
is the evaporation lifetime. Numerically,
\begin{equation}
    \tau_\star
    \simeq
    4.14\times10^{-10}\,{\rm s}
    \left(
        \frac{M_\star}{10^6\,{\rm g}}
    \right)^3,
    \label{eq:PBH_lifetime_numerical}
\end{equation}
in agreement with Refs.~\cite{Vanvlasselaer:2026vkh,Klipfel:2026nzx}. The strong increase of the
Hawking emission rate during the final stage of evaporation is central to the
PBH explosion baryogenesis mechanism discussed below.

The comoving PBH number density remains constant prior to complete
evaporation. The corresponding energy released via Hawking radiation
per unit comoving volume and per unit time is therefore
\begin{equation}
    {\cal E}^{\rm co}(\Delta t)
    \equiv
    n_{\rm PBH,\star}^{\rm co}
    \left(-\frac{dM}{dt}\right)
    =
    n_{\rm PBH,\star}^{\rm co}
    \frac{\mathcal{A}_f}{M^2(\Delta t)}.
    \label{eq:comoving_energy_injection}
\end{equation}
This quantity describes the continuous transfer of PBH mass into 
Hawking radiation. Its contribution to the evolution of the radiation
background and entropy is discussed in Sec.~\ref{subsec:thermal_history},
while the directly emitted graviton component is treated separately in
Sec.~\ref{subsec:hawking_gravitons}.

\subsection{Radiation and entropy evolution}
\label{subsec:thermal_history}
Particles emitted via Hawking radiation can deposit
energy into the surrounding plasma, thereby modifying
both the expansion history and the entropy density
of the Universe. Following
Refs.~\cite{Hooper:2019gtx,Cheek:2021odj}, we describe this evolution directly
in terms of the physical PBH and radiation energy densities.

For the monochromatic PBH population considered here, the PBH energy density
evolves according to
\begin{equation}
    \dot{\rho}_{\rm PBH}
    +
    3H\rho_{\rm PBH}
    =
    \frac{\dot M}{M}\rho_{\rm PBH},
    \label{eq:PBH_energy_evolution}
\end{equation}
where the term $3H\rho_{\rm PBH}$ accounts for the dilution due to cosmic
expansion, while the right-hand side describes the decrease of the PBH energy
density associated with Hawking evaporation. Using
Eq.~\eqref{eq:hawking_mass_loss}, this equation becomes
\begin{equation}
    \dot{\rho}_{\rm PBH}
    +
    3H\rho_{\rm PBH}
    =
    -
    \frac{\mathcal{A}_f}{M^3(t)}
    \rho_{\rm PBH}.
    \label{eq:PBH_energy_evolution_Af}
\end{equation}
The energy lost by the PBHs is injected into the surrounding radiation bath.
The corresponding radiation energy density therefore satisfies~\cite{Hooper:2019gtx,Cheek:2021odj}
\begin{equation}
    \dot{\rho}_{\rm rad}
    +
    4H\rho_{\rm rad}
    =
    -
    \frac{\dot M}{M}\rho_{\rm PBH}
    =
    \frac{\mathcal{A}_f}{M^3(t)}
    \rho_{\rm PBH}.
    \label{eq:radiation_energy_evolution}
\end{equation}
The Hubble rate is determined self-consistently by the total energy density,
\begin{equation}
    H^2
    =
    \frac{1}{3M_{\rm Pl}^2}
    \left(
        \rho_{\rm rad}
        +
        \rho_{\rm PBH}
    \right),
    \qquad
    \frac{\dot a}{a}=H.
    \label{eq:Friedmann_PBH_radiation}
\end{equation}
We take the horizon entry of the characteristic bubble scale
as the starting point of the evaporation calculation, with
the scale factor given by $a_{H,\star}=H_i\chi_i$.
The physical PBH energy density at this epoch is therefore
\begin{equation}
    \left.
    \rho_{\rm PBH}
    \right|_{a=a_{H,\star}}
    =
    \frac{
        n_{\rm PBH,\star}^{\rm co}M_\star
    }{
        a_{H,\star}^3
    }.
    \label{eq:PBH_initial_density}
\end{equation}
The radiation energy density at horizon
entry is
\begin{equation}
    \left.
    \rho_{\rm rad}
    \right|_{a=a_{H,\star}}
    \simeq
    3M_{\rm Pl}^2H_{\rm form,\star}^2,
    \label{eq:radiation_initial_density}
\end{equation}
where the last approximation applies for
$\kappa_H^{\rm SB}\ll1$. The Universe is therefore radiation dominated at
horizon entry. Since the PBH energy density redshifts more slowly than
radiation, however, the PBH component can become dynamically important before
evaporation.
Assuming that the Standard Model particles emitted by the PBHs thermalize
efficiently with the ambient plasma, the background temperature is determined
by
\begin{equation}
    \rho_{\rm rad}(T_b)
    =
    \frac{\pi^2}{30}
    g_*(T_b)T_b^4,
    \label{eq:radiation_temperature}
\end{equation}
and the entropy density is
\begin{equation}
    s_{\rm ent}(T_b)
    =
    \frac{2\pi^2}{45}
    g_{*s}(T_b)T_b^3.
    \label{eq:entropy_density}
\end{equation}
We use $s_{\rm ent}$ to denote the entropy density
in order to distinguish it from the spectator
scalar field $s$. Here $g_*(T_b)$ and $g_{*s}(T_b)$ denote the
effective relativistic degrees of freedom for energy
and entropy densities, respectively. Following Ref.~\cite{Klipfel:2026nzx}, for simplicity we approximate the
effective relativistic degrees of freedom as constant throughout the numerical
evolution relevant for baryogenesis and take
\begin{equation}
    g_*(T_b)\simeq g_{*s}(T_b)\simeq 106.75.
    \label{eq:gstar_constant_approx}
\end{equation}

For the entropy evolution, we define the comoving
entropy density as
$s_{\rm ent}^{\rm co}\equiv a^3s_{\rm ent}$,
which evolves according to~\cite{Scherrer:1984fd,Klipfel:2026nzx}
\begin{equation}
    \frac{d s_{\rm ent}^{\rm co}}{dt}
    =
    \frac{{\cal E}^{\rm co}}{T_b}.
    \label{eq:comoving_entropy_evolution}
\end{equation}
Together with Eqs.~\eqref{eq:PBH_energy_evolution_Af},
\eqref{eq:radiation_energy_evolution}, and
\eqref{eq:Friedmann_PBH_radiation}, these equations determine the radiation
and entropy evolution throughout the PBH lifetime. We denote the asymptotic comoving entropy density
after complete PBH evaporation by
$s_{{\rm ent},\max}^{\rm co}$.

Whether the PBH population becomes dynamically dominant before evaporation
depends on its initial abundance and lifetime. We discuss this condition in
Sec.~\ref{subsec:density_fluctuations}, where it determines the presence of the GW
contribution sourced by PBH density fluctuations.

\section{Electroweak symmetry restoration and baryogenesis from Hawking radiation of light PBHs}
\label{sec:baryogenesis}

We now estimate the baryon asymmetry that can be produced when
the light PBHs originating from the spectator sector reach
the final stage of Hawking evaporation. The microscopic
thermalization, shock propagation, and CP-violating transport
are described using the results of
Refs.~\cite{Vanvlasselaer:2026vkh,Klipfel:2026nzx};
we do not perform an independent hydrodynamic or electroweak
transport simulation. Our calculation instead supplies the PBH population from the
inflationary bubble source and follows its cosmological
evolution. In particular, $M_\star$ and
$n_{\rm PBH,\star}^{\rm co}$ are determined by the nucleation
and growth history, while $T_b(\Delta t)$ and
$s_{{\rm ent},\max}^{\rm co}$ are obtained from
Sec.~\ref{subsec:thermal_history}.

\subsection{Landau--Pomeranchuk--Migdal thermalization and the explosive stage}
\label{subsec:lpm_explosion}

During the final stage of PBH evaporation,
the thermalization of the energetic QCD component
of Hawking radiation is governed by the
Landau--Pomeranchuk--Migdal (LPM) splitting rate
in the surrounding plasma~\cite{Klipfel:2026nzx}. We evaluate the strong coupling using the one-loop QCD running~\cite{ParticleDataGroup:2024cfk,Deur:2016tte,Kara:2013dua,Prosperi:2006hx},
\begin{equation}
    \alpha_s(T_b)
    =
    \left[
        \alpha_s(m_Z)^{-1}
        +
        \frac{\beta_0}{2\pi}
        \ln\left(\frac{T_b}{m_Z}\right)
    \right]^{-1},
    \qquad
    \beta_0
    =
    11-\frac{2N_f}{3}.
    \label{eq:alpha_s_running}
\end{equation}
For definiteness, we adopt $N_f=6$ as a common benchmark throughout the
temperature range considered. This choice is intended as a simplified
estimate for the LPM thermalization calculation rather than a
threshold-matched determination of the QCD running coupling.
The corresponding LPM thermalization length is taken to be~\cite{Vanvlasselaer:2026vkh,Klipfel:2026nzx}
\begin{equation}
    L_{\rm LPM}(M,T_b)
    =
    \frac{1}{\alpha_s^2(T_b)T_b}
    \left[
        \frac{
            6\beta_{1/2}T_H(M)
        }{
            \pi C_R g_c T_b
            \ln\!\left[\beta_{1/2}T_H(M)/T_b\right]
        }
    \right]^{1/2},
    \label{eq:LPM_length}
\end{equation}
where
\begin{equation}
    \beta_{1/2}=4.53,
    \qquad
    C_R=C_F=\frac{4}{3},
    \qquad
    g_c=2N_c+N_f=12,
    \qquad
    N_c=3.
\end{equation}
Here $\beta_{1/2}T_H$ characterizes the energy near the peak of the primary
spin-$1/2$ Hawking spectrum, while $C_R=C_F$ corresponds to the quark-dominated
color-charged component of the emission.

The final explosive stage begins when the remaining PBH lifetime becomes
comparable to the thermalization scale of the emitted particles. In natural
units, the threshold mass is therefore defined by~\cite{Vanvlasselaer:2026vkh,Klipfel:2026nzx}
\begin{equation}
    \tau\!\left[M_{\rm thres}(T_b)\right]
    =
    L_{\rm LPM}\!\left[M_{\rm thres}(T_b),T_b\right],
    \label{eq:Mthres_implicit}
\end{equation}
which, using Eq.~\eqref{eq:PBH_lifetime}, becomes
\begin{equation}
    \frac{M_{\rm thres}^3(T_b)}{3\mathcal{A}_f}
    =
    L_{\rm LPM}\!\left[M_{\rm thres}(T_b),T_b\right].
    \label{eq:Mthres_equation}
\end{equation}
For each background temperature $T_b$,
Eq.~\eqref{eq:Mthres_equation} is solved numerically
for the threshold mass $M_{\rm thres}$.

For the monochromatic population, all PBHs reach the threshold at the same
time. With $\Delta t=0$ defined at horizon entry as in
Sec.~\ref{subsec:hawking_evaporation}, the explosion time is determined
implicitly by
\begin{equation}
    M_\star^3
    -
    3\mathcal{A}_f\Delta t_{\rm expl}
    =
    M_{\rm thres}^3
    \!\left[T_b(\Delta t_{\rm expl})\right].
    \label{eq:threshold_explosion_time}
\end{equation}

Since the initial PBH population is monochromatic, the comoving explosion rate
is a single delta source,
\begin{equation}
    \Gamma_{\rm expl}^{\rm co,SB}(\Delta t)
    =
    n_{\rm PBH,\star}^{\rm co}
    \delta\!\left(\Delta t-\Delta t_{\rm expl}\right).
    \label{eq:single_mass_explosion_rate}
\end{equation}

\subsection{Local electroweak symmetry restoration}
\label{subsec:ew_restoration}

When a PBH reaches $M_{\rm thres}$, its remaining energy is deposited over a
region whose characteristic size is $L_{\rm LPM}$. Following the hydrodynamic
treatment of Ref.~\cite{Klipfel:2026nzx}, the effect of the preceding
non-instantaneous heating stage is encoded by a factor $K$. We adopt the
benchmark value $K=5$~\cite{Vanvlasselaer:2026vkh}, which is representative of the hydrodynamic simulations
used in Ref.~\cite{Klipfel:2026nzx}. The initial fireball temperature is then estimated
from the ambient radiation energy density and the deposited energy as
\begin{equation}
    T_{\rm fb}^4(T_b,K)
    =
    T_b^4
    +
    \frac{
        90K M_{\rm thres}(T_b)
    }{
        4\pi^3 g_{*,\rm EW}
        L_{\rm LPM}^3[M_{\rm thres}(T_b),T_b]
    },
    \label{eq:fireball_temperature}
\end{equation}
where $g_{*,\rm EW}=106.75$. We use the electroweak restoration temperature
$T_{\rm EW}=162\,{\rm GeV}$, following Ref.~\cite{Vanvlasselaer:2026vkh,Klipfel:2026nzx}.

The minimum background temperature for which the explosion can still restore
electroweak symmetry locally is defined by
\begin{equation}
    T_{\rm fb}(T_{\min},K)
    =
    T_{\rm EW}.
    \label{eq:Tmin_definition}
\end{equation}
For $T_b<T_{\min}$, the final PBH explosion no longer heats the local plasma
above $T_{\rm EW}$ and therefore does not contribute to the baryogenesis
mechanism considered here.

The subsequent hydrodynamic evolution produces a finite region whose
temperature exceeds the electroweak restoration scale. Following
Refs.~\cite{Vanvlasselaer:2026vkh,Klipfel:2026nzx}, we characterize this
region by its maximum radius $R_{\max}^{\rm EW}$, which determines the
spatial volume relevant for the baryogenesis calculation below. The maximum radius reached by the electroweak-symmetric
shell is parameterized by
\begin{equation}
    R_{\max}^{\rm EW}(T_b)
    =
    0.84
    \left[
        \frac{
            K M_{\rm thres}(T_b)
        }{
            p(T_{\rm EW})
        }
    \right]^{1/3},
    \label{eq:Rmax_EW}
\end{equation}
with
\begin{equation}
    p(T_{\rm EW})
    =
    \frac{\pi^2}{90}
    g_{*,\rm EW}T_{\rm EW}^4.
    \label{eq:pressure_EW}
\end{equation}
The moving boundaries of this shell provide the out-of-equilibrium interfaces
at which the local baryon asymmetry is generated.

\subsection{Sphaleron window for baryogenesis}
\label{subsec:baryogenesis_window}

Baryogenesis is operative only when sphaleron processes in the ambient plasma
are inefficient; otherwise the generated baryon asymmetry is washed out. We
adopt the following parametrization for the sphaleron rate normalized by the
temperature~\cite{DOnofrio:2014rug,Hong:2023zrf,Mariotti:2024eoh,Klipfel:2026nzx},
\begin{equation}
    \frac{\Gamma_{\rm spha}(T)}{T}
    =
    \begin{cases}
        8.0\times10^{-7},
        & T>T_{\rm EW},\\[4pt]
        e^{-147.7+0.83\left(T/{\rm GeV}\right)},
        & T\leq T_{\rm EW}.
    \end{cases}
    \label{eq:sphaleron_rate}
\end{equation}
At a given time, the sphaleron freeze-out temperature is determined from the
instantaneous expansion rate through
\begin{equation}
    \Gamma_{\rm spha}
    \!\left[T_{\rm sph}(\Delta t)\right]
    =
    H(\Delta t).
    \label{eq:sphaleron_freezeout}
\end{equation}
The lower end of the baryogenesis temperature window is set by the loss of
local electroweak restoration and is given by $T_{\min}$. For the
monochromatic source, successful baryogenesis requires the background
temperature at the explosion time to satisfy
\begin{equation}
    T_{\min}
    <
    T_b(\Delta t_{\rm expl})
    <
    T_{\rm sph}(\Delta t_{\rm expl}).
    \label{eq:baryogenesis_time_window}
\end{equation}
It is convenient to encode this condition by
\begin{equation}
    {\cal I}_{\rm BG}(\Delta t)
    \equiv
    \Theta\!\left[T_b(\Delta t)-T_{\min}\right]
    \Theta\!\left[T_{\rm sph}(\Delta t)-T_b(\Delta t)\right].
    \label{eq:baryogenesis_window_function}
\end{equation}
This form imposes the baryogenesis window directly in temperature space and
does not require the background temperature to evolve monotonically.

The single explosion calculation also assumes that neighboring
hydrodynamic shells do not overlap appreciably before the baryogenesis process
is completed. Following the validity condition of Ref.~\cite{Klipfel:2026nzx},
the characteristic physical PBH separation at the explosion time,
\begin{equation}
    d_{\rm PBH}(\Delta t_{\rm expl})
    =
    a(\Delta t_{\rm expl})
    \left(n_{\rm PBH,\star}^{\rm co}\right)^{-1/3},
    \label{eq:PBH_separation_explosion}
\end{equation}
is required to satisfy
\begin{equation}
    d_{\rm PBH}(\Delta t_{\rm expl})
    >
    R_{\max}^{\rm EW}
    \!\left[T_b(\Delta t_{\rm expl})\right].
    \label{eq:no_shock_overlap}
\end{equation}
This condition ensures that the isolated-shock baryon yield used below remains
a consistent approximation.

\subsection{Baryon asymmetry}
\label{subsec:baryon_asymmetry}
We now relate the baryon number generated by an individual PBH explosion to the resulting cosmological baryon asymmetry. In the baryogenesis sector, we adopt the effective dimension-five CP-violating interaction considered in Refs.~\cite{Klipfel:2026nzx,Bodeker:2020ghk,Barni:2025ifb,Espinosa:2011eu} and make use of the corresponding numerical results from their transport analysis,
\begin{equation}
    {\cal L}_{\rm CPV}
    \supset
    -\frac{{\cal S}}{\Lambda}\,
    \overline{Q}_3\,\widetilde H\,
    \left(
        a_{\rm CPV}
        +
        i b_{\rm CPV}\gamma_5
    \right)
    t_R
    +{\rm h.c.},
    \label{eq:CPV_operator}
\end{equation}
where \(Q_3=(t_L,b_L)^T\) denotes the third-generation left-handed quark doublet, \(t_R\) the right-handed top quark, \(\widetilde H=i\sigma_2H^*\), and \({\cal S}\) an additional scalar field associated with the CP-violating sector. The scale \(\Lambda\) characterizes the higher-dimensional interaction, while \(b_{\rm CPV}\) controls its CP-violating component. We emphasize that \({\cal S}\) is introduced solely as part of the effective baryogenesis sector and is distinct from the spectator field \(s\) responsible for bubble nucleation in our construction.

For the benchmark wall profile and transport calculation considered in
Ref.~\cite{Klipfel:2026nzx}, the local baryon asymmetry generated at the
moving interface can be parameterized as 
\begin{equation}
    \frac{\eta_B^{\rm loc}}{\eta_B^{\rm obs}}
    \simeq
    10\,b_{\rm CPV}
    \left(\frac{1\,{\rm TeV}}{\Lambda}\right)^2,
    \label{eq:local_BAU_CPV}
\end{equation}
where
\[
    \eta_B^{\rm obs}\equiv \frac{n_B}{s_{\rm ent}}
    =8.65\times10^{-11}
\]
denotes the observed baryon-to-entropy ratio~\cite{Bodeker:2020ghk,vandeVis:2025efm,Cooke:2013cba,
Planck:2015fie,Pitrou:2018cgg,Klipfel:2026nzx}.
It is convenient to define
\begin{equation}
    C_{\rm CPV}
    \equiv
    \frac{\eta_B^{\rm loc}}{\eta_B^{\rm obs}}
    \left(\frac{\Lambda}{{\rm TeV}}\right)^2,
    \label{eq:CCPV_definition}
\end{equation}
such that the benchmark adopted below corresponds to
$C_{\rm CPV}\simeq 10\,b_{\rm CPV}$, with $b_{\rm CPV}=1$.

For an individual PBH exploding in a plasma with background
temperature $T_b$, we estimate the resulting baryon number
by integrating over the region traversed by the
electroweak-restoring interface~\cite{Klipfel:2026nzx},
\begin{equation}
    N_B^{(1)}(T_b)
    =
    \eta_B^{\rm loc}
    s_{\rm ent}(T_{\rm EW})
    \frac{4\pi}{3}
    \left[R_{\max}^{\rm EW}(T_b)\right]^3.
    \label{eq:single_explosion_baryon_number}
\end{equation}
Here, $\eta_B^{\rm loc}s_{\rm ent}(T_{\rm EW})$ gives the baryon number density
generated locally as the electroweak-restoring interface propagates through
the plasma, while the factor $4\pi [R_{\max}^{\rm EW}]^3/3$ represents the
total volume swept out by the interface.

Approximating the plasma near the electroweak scale as a relativistic
radiation fluid, one has
\begin{equation}
    s_{\rm ent}(T_{\rm EW})
    \simeq
    \frac{4p(T_{\rm EW})}{T_{\rm EW}}.
\end{equation}
Together with Eqs.~\eqref{eq:Rmax_EW} and \eqref{eq:pressure_EW}, this gives
\begin{equation}
    N_B^{(1)}(T_b)
    =
    \eta_B^{\rm loc}
    (0.84)^3
    \frac{16\pi}{3}
    \frac{K M_{\rm thres}(T_b)}{T_{\rm EW}}.
    \label{eq:single_explosion_baryon_number_simplified}
\end{equation}
Using Eq.~\eqref{eq:CCPV_definition}, the same result can be written as
\begin{equation}
    N_B^{(1)}(T_b)
    =
    \eta_B^{\rm obs}
    C_{\rm CPV}
    \left(\frac{{\rm TeV}}{\Lambda}\right)^2
    (0.84)^3
    \frac{16\pi}{3}
    \frac{K M_{\rm thres}(T_b)}{T_{\rm EW}}.
    \label{eq:single_explosion_baryon_number_CCPV}
\end{equation}

The cumulative baryon number per unit comoving volume is then
\begin{equation}
    n_B^{\rm co}
    =
    \int d\Delta t\,
    N_B^{(1)}[T_b(\Delta t)]
    \Gamma_{\rm expl}^{\rm co,SB}(\Delta t)
    {\cal I}_{\rm BG}(\Delta t),
    \label{eq:comoving_baryon_number}
\end{equation}
where $\Gamma_{\rm expl}^{\rm co,SB}$ is the explosion rate per
unit comoving volume and ${\cal I}_{\rm BG}$ implements the conditions under
which shock baryogenesis is operative.

For the monochromatic source in
Eq.~\eqref{eq:single_mass_explosion_rate}, all PBHs undergo their final
explosive stage at the same time, and
Eq.~\eqref{eq:comoving_baryon_number} reduces to
\begin{equation}
    n_B^{\rm co}
    =
    n_{\rm PBH,\star}^{\rm co}
    N_B^{(1)}
    \!\left[T_b(\Delta t_{\rm expl})\right]
    {\cal I}_{\rm BG}(\Delta t_{\rm expl}).
    \label{eq:comoving_baryon_number_single_mass}
\end{equation}
For a fixed baryon yield per explosion and within the
baryogenesis window, the total comoving baryon number scales
linearly with the comoving PBH abundance.
The final baryon-to-entropy ratio, however, also depends
on the entropy generated during PBH evaporation,
as discussed below.

To compare the generated asymmetry with the observed value, we define
\begin{equation}
    Y_{\rm SB}
    \equiv
    \frac{\eta_{B,\rm SB}^{\rm tot}}{\eta_B^{\rm obs}},
    \qquad
    Y_{\rm scaled}^{\rm SB}
    \equiv
    Y_{\rm SB}
    \left(\frac{\Lambda}{{\rm TeV}}\right)^2.
    \label{eq:Yscaled_definition}
\end{equation}

The thermal evolution discussed in
Sec.~\ref{subsec:thermal_history} determines the comoving entropy density
$s_{\rm ent}^{\rm co}=a^3s_{\rm ent}$. During PBH evaporation, entropy injection causes
$s_{\rm ent}^{\rm co}$ to increase, while after complete evaporation it approaches
the asymptotic value $s_{{\rm ent},\max}^{\rm co}$ and remains conserved thereafter.
Since $n_B^{\rm co}$ and $s_{{\rm ent},\max}^{\rm co}$ are defined using the same
comoving-volume normalization, the final baryon-to-entropy ratio is simply
\begin{equation}
    \eta_{B,\rm SB}^{\rm tot}
    =
    \frac{n_B^{\rm co}}{s_{{\rm ent},\max}^{\rm co}}.
    \label{eq:etaB_total_definition}
\end{equation}

Combining Eqs.~\eqref{eq:single_explosion_baryon_number_CCPV},
\eqref{eq:comoving_baryon_number},
\eqref{eq:Yscaled_definition}, and
\eqref{eq:etaB_total_definition}, the scaled cosmological yield becomes
\begin{equation}
    Y_{\rm scaled}^{\rm SB}
    =
    {\cal Q}_{\rm phys}^{\rm SB}
    \int d\Delta t\,
    M_{\rm thres}[T_b(\Delta t)]
    \Gamma_{\rm expl}^{\rm co,SB}(\Delta t)
    {\cal I}_{\rm BG}(\Delta t),
    \label{eq:Yscaled_integral}
\end{equation}
with
\begin{equation}
    {\cal Q}_{\rm phys}^{\rm SB}
    =
    C_{\rm CPV}
    (0.84)^3
    \frac{16\pi}{3}
    \frac{K}{T_{\rm EW}}
    \frac{1}{s_{{\rm ent},\max}^{\rm co}}.
    \label{eq:BAU_prefactor}
\end{equation}
Thus, all entropy dilution associated with the PBH evaporation history is
encoded directly in the asymptotic comoving entropy density
$s_{{\rm ent},\max}^{\rm co}$.

Finally, substituting the monochromatic explosion rate gives
\begin{align}
    Y_{\rm scaled}^{\rm SB}
    &=
    {\cal Q}_{\rm phys}^{\rm SB}
    n_{\rm PBH,\star}^{\rm co}
    M_{\rm thres}
    \!\left[T_b(\Delta t_{\rm expl})\right]
    \nonumber\\
    &\quad\times
    \Theta\!\left[
        T_b(\Delta t_{\rm expl})-T_{\min}
    \right]
    \Theta\!\left[
        T_{\rm sph}(\Delta t_{\rm expl})
        -T_b(\Delta t_{\rm expl})
    \right].
    \label{eq:Yscaled_single_mass_final}
\end{align}
The final baryon asymmetry is therefore controlled by both the comoving PBH
abundance and the entropy generated during PBH evaporation, together with the
thermal and hydrodynamic conditions governing the final explosive stage. In
particular, once the PBH abundance is sufficiently large for a PBH dominated
epoch to develop, increasing the PBH population also increases the entropy
injected during evaporation. As found in Ref.~\cite{Klipfel:2026nzx}, these
two effects can approximately compensate each other, causing the final baryon
yield to approach a saturation value rather than continuing to grow linearly
with the PBH abundance.

\section{GW signatures}
\label{sec:gravitational_waves}

The PBH population generated by the spectator sector transition gives rise to
two qualitatively distinct GW contributions. First, gravitons are emitted
directly through Hawking radiation from the PBHs. This contribution is present
regardless of whether the PBHs ever dominate the cosmic energy density.
Second, if the PBH abundance is sufficiently large to produce a transient
PBH dominated epoch before evaporation, fluctuations in the PBH number density
source scalar perturbations that subsequently induce GWs at second order.
These two production mechanisms are illustrated
schematically in Fig.~\ref{fig:PBH_mechanism}(e)
and (c), respectively, and are discussed
separately below~\cite{Ireland:2023avg,Dolgov:2011cq,
Domenech:2020ssp,Domenech:2024wao}.

\subsection{GWs from Hawking radiation}
\label{subsec:hawking_gravitons}

For the nonrotating PBHs considered here, the instantaneous graviton energy
spectrum can be written as~\cite{Page:1976df,Ireland:2023avg}
\begin{equation}
    \frac{d^2E_g}{dt\,d\omega}
    =
    \frac{1}{2\pi}
    \sum_{\ell,m}
    \frac{
        \omega\,\Gamma_{\ell m}^{(2)}(\omega,M)
    }{
        e^{\omega/T_H(M)}-1
    },
    \label{eq:hawking_graviton_spectrum}
\end{equation}
where $\Gamma_{\ell m}^{(2)}$ denotes the spin-2 greybody factor, including the
graviton polarization multiplicity, and
$T_H(M)=M_{\rm Pl}^2/M$ is the Hawking temperature in the reduced Planck mass
convention adopted throughout this work. The greybody factor encodes the
propagation of the emitted gravitons through the Schwarzschild geometry and
therefore modifies the emission spectrum relative to that of an ideal
blackbody~\cite{Ireland:2023avg}.

As the PBH evaporates, its mass decreases according to
Eq.~\eqref{eq:PBH_mass_evolution}, while its Hawking temperature increases.
The graviton spectrum observed today therefore receives contributions from the
entire evaporation history, with the emitted frequencies subsequently redshifted
by the cosmological expansion. A complete calculation can be performed by
integrating Eq.~\eqref{eq:hawking_graviton_spectrum} over the evolving PBH mass
and background history, as discussed in Ref.~\cite{Ireland:2023avg}.

For the numerical evaluation of the GW contribution
from Hawking radiation, we recast the semi-analytic estimate of Ref.~\cite{Borah:2026zbl} in terms of the PBH
parameters obtained in our scenario. It is convenient to separate the overall
normalization from the frequency dependence and write
\begin{equation}
    \Omega_{\rm GW,0}^{\rm Haw}(f_0)h^2
    =
    {\cal A}_{\rm Haw}
    \left(
        \frac{f_0}{10^9\,{\rm Hz}}
    \right)^4
    {\cal F}_{\rm Haw}
    \left(
        x_0;z_{\max}^{\rm SB}
    \right),
    \label{eq:Omega_GW_Hawking_template}
\end{equation}
where
\begin{equation}
    {\cal A}_{\rm Haw}
    =
    1.36\times10^{-21}
    \left(
        \frac{g_{\ast,H,\star}}{100}
    \right)^2
    \left(
        \frac{1\,{\rm g}}{M_\star}
    \right)^2,
    \label{eq:Hawking_GW_amplitude}
\end{equation}
and
\begin{equation}
    x_0
    \equiv
    \frac{2\pi f_0}{T_{\rm BH,0}^{\rm SB}}.
    \label{eq:Hawking_dimensionless_frequency}
\end{equation}
Here
$g_{\ast,H,\star}\equiv g_{\ast,H}[T_H(M_\star)]$
denotes the effective number of particle degrees of freedom contributing to
Hawking evaporation at the initial PBH temperature.
The quantity $T_{\rm BH,0}^{\rm SB}$ characterizes the Hawking temperature
redshifted to the present epoch and expressed in frequency units. For the
monochromatic PBH population considered here, it is approximated by
\begin{equation}
    T_{\rm BH,0}^{\rm SB}
    \simeq
    4.53\times10^{15}\,{\rm Hz}
       \left(
        \frac{M_\star}
             {10^5\,{\rm g}}
    \right)^{1/2}
    \left[
        \frac{100}
             {g_{\ast s}(T_{\rm ev})}
    \right]^{1/12}
    \left(
        \frac{100}
             {g_{\ast,H,\star}}
    \right)^{1/2},
    \label{eq:TBH0_Hawking_template}
\end{equation}
where $T_{\rm ev}$ is the background temperature at the completion of PBH
evaporation. The factor $2\pi$ in Eq.~\eqref{eq:Hawking_dimensionless_frequency}
accounts for the conversion between the ordinary frequency $f_0$ and the
corresponding angular frequency.
The spectral shape is encoded in
\begin{equation}
    {\cal F}_{\rm Haw}
    \left(
        x_0;z_{\max}^{\rm SB}
    \right)
    =
    \int_{1}^{1+z_{\max}^{\rm SB}}
    dy\,
    \frac{
        y^{1/2}
    }{
        e^{x_0 y}-1
    },
    \label{eq:I_Hawking_template}
\end{equation}
where we have introduced $y=1+z$. The upper integration limit is determined
by
\begin{equation}
    1+z_{\max}^{\rm SB}
    \simeq
    7.36\times10^7
    \left(
        \frac{M_\star}
             {1\,{\rm g}}
    \right)^{4/3}
    \left(
        \kappa_H^{\rm SB}
    \right)^{1/3}.
    \label{eq:zmax_Hawking_template}
\end{equation}
Here $\kappa_H^{\rm SB}$ denotes the PBH abundance at horizon entry, as defined
in Eq.~\eqref{eq:PBH_fraction_formation}. Since the PBH energy density at
formation is much smaller than the radiation energy density in the parameter
region considered here, the PBH-to-radiation ratio
$\rho_{\rm PBH}/\rho_{\rm rad}$ and the PBH fraction of the total energy density
$\rho_{\rm PBH}/\rho_{\rm tot}$ are numerically equivalent to the accuracy
relevant for this work. We therefore do not distinguish between the two
conventions and denote both by $\kappa_H^{\rm SB}$.

The GW signal from direct graviton emission is a generic
consequence of PBH evaporation. The semi-analytic spectrum above is adopted for the benchmark
cosmological history in which the PBH population undergoes an early
PBH dominated epoch before evaporation~\cite{Borah:2026zbl}. The condition for
realizing this epoch is discussed in Sec.~\ref{subsec:density_fluctuations}.

\subsection{GWs induced by PBH density fluctuations}
\label{subsec:density_fluctuations}

We now consider the GW background induced by PBH density
fluctuations in the PBH-dominated regime.
Before deriving the corresponding spectrum, we specify
the conditions for PBH domination and consistency with
big bang nucleosynthesis (BBN).

In the PBH dominated branch, the temperature of the radiation bath after
evaporation can be estimated as~\cite{Domenech:2020ssp}
\begin{equation}
    T_{\rm ev}
    \simeq
    2.70\times10^4\,{\rm GeV}
    \left(
        \frac{M_\star}{10^4\,{\rm g}}
    \right)^{-3/2}
    \left[
        \frac{g_{\ast,H,\star}}{100}
    \right]^{1/2}
    \left[
        \frac{g_*(T_{\rm ev})}{100}
    \right]^{-1/4},
    \label{eq:PBH_evaporation_temperature}
\end{equation}
Requiring the PBHs to evaporate sufficiently early for
successful BBN, we impose the conservative condition
$T_{\rm ev}>4\,{\rm MeV}$, which gives the approximate
upper bound $M_\star\lesssim4\times10^8\,{\rm g}$.

For the PBH dominated branch considered here, the PBHs must become the
dominant component before evaporation. During radiation domination, the PBH
energy density redshifts as $a^{-3}$, whereas the radiation energy density
redshifts as $a^{-4}$, so that the PBH fraction grows with the scale factor.
Requiring PBH domination to occur before evaporation therefore imposes a lower
bound on the initial PBH abundance~\cite{Domenech:2020ssp}. With the horizon-mass convention adopted
in this work, we use
\begin{equation}
    \kappa_H^{\rm SB}
    \gtrsim
    \kappa_{\min}^{\rm SB},
    \qquad
    \kappa_{\min}^{\rm SB}
    \simeq
    2.84\times10^{-10}
    \left(
        \frac{M_\star}{10^4\,{\rm g}}
    \right)^{-1}
    \left[
        \frac{g_{\ast,H,\star}}{100}
    \right]^{1/2}.
    \label{eq:beta_min_SB}
\end{equation}

For a monochromatic and initially unclustered PBH population that undergoes an
early PBH dominated epoch, Poisson density fluctuations generate an additional
GW background. The requirement that this induced GW contribution remain
consistent with BBN places an upper bound on the initial PBH abundance~\cite{Domenech:2020ssp}. In the mass convention adopted here, this bound is
\begin{equation}
    \kappa_H^{\rm SB}
    \lesssim
    \kappa_{\max}^{\rm GW/BBN},
    \label{eq:beta_max_GW_BBN_condition}
\end{equation}
with
\begin{equation}
    \kappa_{\max}^{\rm GW/BBN}
    \simeq
    4.92\times10^{-7}
        \left(
        \frac{M_\star}{10^4\,{\rm g}}
    \right)^{-17/24}
    \left[
        \frac{g_{\ast,H,\star}}{100}
    \right]^{17/48}
    \left[
        \frac{g_*(T_{\rm ev})}{100}
    \right]^{1/16}.
    \label{eq:beta_max_GW_BBN}
\end{equation}
This constraint assumes a narrow PBH mass distribution, approximately Poisson
spatial statistics, a PBH dominated epoch before evaporation, and the
sudden-evaporation treatment used in Ref.~\cite{Domenech:2020ssp}.

The parameter region relevant for the density fluctuation GW signal therefore
satisfies
\begin{equation}
    \kappa_{\min}^{\rm SB}
    \lesssim
    \kappa_H^{\rm SB}
    \lesssim
    \kappa_{\max}^{\rm GW/BBN},
    \label{eq:beta_allowed_GW}
\end{equation}
together with the pre-BBN evaporation condition. For the benchmark values
$g_{\ast,H,\star}\simeq100$ and $g_*(T_{\rm ev})\simeq100$, the corresponding
abundance interval becomes
\begin{equation}
    2.84\times10^{-10}
    \left(
        \frac{M_\star}{10^4\,{\rm g}}
    \right)^{-1}
    \lesssim
    \kappa_H^{\rm SB}
    \lesssim
    4.92\times10^{-7}
    \left(
        \frac{M_\star}{10^4\,{\rm g}}
    \right)^{-17/24}.
    \label{eq:beta_allowed_GW_benchmark}
\end{equation}

We now turn to the associated GW spectrum. Even in the absence of primordial
clustering, the discrete nature of the PBH population generates Poisson
fluctuations in its number density. These fluctuations initially correspond to
isocurvature perturbations. During a PBH dominated epoch, they source scalar
metric perturbations, which subsequently generate GWs at second order. The
production is strongly enhanced around the transition from PBH domination back
to radiation domination at evaporation~\cite{Domenech:2020ssp}.

Following the semi-analytic template used in
Ref.~\cite{Borah:2026zbl}, we approximate the present-day spectrum as
\begin{equation}
    \Omega_{\rm GW,0}^{\rm DF}(f_0)h^2
    \simeq
    \Omega_{\rm GW,0}^{\rm DF,peak}h^2
    \left(
        \frac{f_0}
             {f_{\rm peak}^{\rm DF}}
    \right)^{11/3}
    \Theta\!\left(
        f_{\rm peak}^{\rm DF}-f_0
    \right).
    \label{eq:Omega_GW_density_template}
\end{equation}
The spectrum therefore rises approximately as $f_0^{11/3}$ below the peak and
is cut off at the ultraviolet scale associated with the mean PBH separation~\cite{Domenech:2020ssp,Borah:2026zbl}.

Expressed directly in terms of the PBH-to-radiation ratio derived in our
scenario, the peak amplitude is
\begin{equation}
    \Omega_{\rm GW,0}^{\rm DF,peak}
    \simeq
    1.46\times10^{-4}
       \left(
        \frac{M_\star}
             {10^7\,{\rm g}}
    \right)^{34/9}
    \left(
        \frac{\kappa_H^{\rm SB}}
             {10^{-8}}
    \right)^{16/3}.
    \label{eq:Omega_GW_density_peak}
\end{equation}
The corresponding peak frequency is
\begin{equation}
    f_{\rm peak}^{\rm DF}
    \simeq
    1.7\times10^3\,{\rm Hz}
    \left(
        \frac{M_\star}
             {10^4\,{\rm g}}
    \right)^{-5/6}.
    \label{eq:fpeak_density_GW}
\end{equation}
Thus lighter PBHs shift the signal toward higher frequencies, whereas the peak
amplitude depends strongly on both the PBH mass and the abundance inherited
from the spectator sector tunneling history.

The template in
Eqs.~\eqref{eq:Omega_GW_density_template}--\eqref{eq:fpeak_density_GW}
is applied only to parameter points that undergo an early PBH dominated epoch.
It further assumes a sufficiently narrow PBH mass distribution and an
approximately Poisson spatial distribution. 

\section{Numerical results}
\label{sec:numerical_results}

We now present the numerical results of the spectator sector scenario developed
in the preceding sections. Throughout this section, we fix
\begin{equation}
    \Delta N_{\rm nuc}=1
\end{equation}
as a common benchmark. As shown in
Eq.~\eqref{eq:bubble_source_peak_width}, the bubble number density depends
linearly on the effective source width $\Delta N_{\rm nuc}$,
whereas its dominant sensitivity arises from
the nucleation rate $\Gamma_\star$, which depends
exponentially on the tunneling action. The abundance can therefore vary
substantially through changes in the spectator potential parameters, making
$\Delta N_{\rm nuc}=1$ a representative choice for the numerical analysis. We further consider two benchmark values of the effective wall
velocity,
\begin{equation}
    v_{\rm eff}=0.3,
    \qquad
    v_{\rm eff}=0.7.
    \label{eq:numerical_veff_benchmarks}
\end{equation}
As discussed in Sec.~\ref{subsec:bubble_evolution}, $v_{\rm eff}$ is treated
as a phenomenological parameter characterizing the accumulated wall
propagation during inflation. A first-principles determination of this
quantity would require solving the time-dependent bubble wall dynamics in the
evolving spectator potential. The benchmark values above are therefore used
to explore the resulting PBH phenomenology within the effective description
adopted in this work.

For each value of $v_{\rm eff}$, we scan the remaining model parameters and
retain only points satisfying the vacuum structure, tunneling, stability of
the bubble interior, supercriticality, PBH formation, and cosmological
consistency conditions discussed in the preceding sections. In addition, we
restrict the numerical scan to the PBH dominated branch, requiring the PBH
abundance to be sufficiently large for the PBHs to dominate the cosmic energy
density before evaporation. This restriction ensures a sufficiently large PBH
population and selects the regime in which the density fluctuation GW signal
can be significantly enhanced.

To illustrate the resulting PBH phenomenology, we present five
BPs  for each value of $v_{\rm eff}$,
covering different PBH masses, abundances, baryon yields,
and GW signals.
The peak amplitude of the density fluctuation GW contribution is
defined as
\begin{equation}
    \Omega_{\rm GW,0}^{\rm DF,peak}h^2
    \equiv
    \max_{f_0}
    \left[
        \Omega_{\rm GW,0}^{\rm DF}(f_0)h^2
    \right].
    \label{eq:GW_benchmark_ranking}
\end{equation}
The properties of these BPs are summarized in
Table~\ref{tab:numerical_benchmarks}. The first five points correspond to
$v_{\rm eff}=0.3$, while the remaining five correspond to
$v_{\rm eff}=0.7$.

To assess the validity of the monochromatic-limit treatment for these
BPs, we further reconstruct the nucleation history across the
corresponding inflationary interval. We define the normalized
number-weighted distribution of the PBH-producing nucleation events as
\begin{equation}
    {\cal P}_N(N)
    \equiv
    \frac{
        e^{-3N}\Gamma(N)/H(N)
    }{
        \displaystyle
        \int_{\rm source} dN'\,
        e^{-3N'}\Gamma(N')/H(N')
    },
    \qquad
    \int_{\rm source} dN\,{\cal P}_N(N)=1.
\end{equation}
To quantify the localization of the nucleation history, we define
$\Delta N_{90}$ as the minimum width of a continuous interval
$[N_-,N_+]$ that contains $90\%$ of the resulting PBH population
by number, namely,
\begin{equation}
    \Delta N_{90}
    \equiv
    \min_{N_-<N_+}
    \left\{
        N_+-N_-
        \;\bigg|\;
        \int_{N_-}^{N_+} dN\,{\cal P}_N(N)\geq 0.9
    \right\},
\end{equation}
where both endpoints are restricted to the source interval.

As a conservative test, we evaluate this quantity at
$\sigma_{N,\max}$, the largest value of $\sigma_N$ found in
our numerical scan for each BP that satisfies the physical
consistency conditions discussed in the preceding sections.
This value can reach order unity, but represents a
phenomenological upper limit within our scan rather than
one derived from solving the bubble wall equation of motion.
Nevertheless, the PBH-producing nucleation histories remain
substantially more localized owing to the exponential
dependence of the tunneling rate on the bounce action. For all ten BPs we find
\begin{equation}
    \Delta N_{90}<1.
\end{equation}
These results demonstrate that the
PBH-producing nucleation history can be substantially
more localized than the underlying spectator potential
feature, even when the latter extends over a relatively
broad interval. This localization motivates the use of
the monochromatic approximation as a leading phenomenological
description in Sec.~\ref{subsec:monochromatic_source}, while
confirming that the physical nucleation width relevant for the PBH
mass distribution is not determined directly by $\sigma_N$.

\begin{table*}[t]
    \centering
    \small
    \setlength{\tabcolsep}{4.5pt}
    \renewcommand{\arraystretch}{1.15}
\caption{
Representative BPs for $\Delta N_{\rm nuc}=1$ and
$v_{\rm eff}=0.3$ or $0.7$. Five BPs are presented for
each value of $v_{\rm eff}$ to illustrate the resulting PBH phenomenology.
The PBH mass, abundance, baryon yield, and GW observables are derived
quantities, while the corresponding spectator potential parameters are
listed in Table~\ref{tab:benchmark_potential_parameters}.
The baryon yield is evaluated for the benchmark $\Lambda=1\,{\rm TeV}$.
}
    \label{tab:numerical_benchmarks}
    \begin{tabular}{c c c c c c c c}
        \hline\hline
        BP
        &
        $v_{\rm eff}$
        &
        $N_\star$
        &
        $M_\star\,[{\rm g}]$
        &
        $\kappa_H^{\rm SB}$
        &
        $Y_{\rm SB}$
        &
        $f_{\rm peak}^{\rm DF}\,[{\rm Hz}]$
        &
        $\Omega_{\rm GW,0}^{\rm DF,peak}h^2$
        \\
        \hline
BP1 & 0.3 & 7.5172 & $4.024\times10^{6}$ & $6.567\times10^{-9}$ & $1.141\times10^{-3}$ & $1.148\times10^{1}$ & $2.26\times10^{-7}$ \\
BP2 & 0.3 & 6.2759 & $3.096\times10^{6}$ & $6.541\times10^{-9}$ & $1.956\times10^{-3}$ & $1.428\times10^{1}$ & $8.215\times10^{-8}$ \\
BP3 & 0.3 & 7.2069 & $2.162\times10^{6}$ & $6.563\times10^{-9}$ & $4.064\times10^{-3}$ & $1.926\times10^{1}$ & $2.157\times10^{-8}$ \\
BP4 & 0.3 & 5.0345 & $2.366\times10^{6}$ & $6.088\times10^{-9}$ & $3.385\times10^{-3}$ & $1.787\times10^{1}$ & $2.028\times10^{-8}$ \\
BP5 & 0.3 & 5.3448 & $4.781\times10^{5}$ & $2.578\times10^{-9}$ & $8.307\times10^{-2}$ & $6.774\times10^{1}$ & $4.939\times10^{-13}$ \\
        \hline
BP6 & 0.7 & 5.6552 & $5.256\times10^{5}$ & $1.235\times10^{-8}$ & $6.891\times10^{-2}$ & $6.26\times10^{1}$ & $2.998\times10^{-9}$ \\
BP7 & 0.7 & 6.5862 & $3.398\times10^{6}$ & $1.713\times10^{-11}$ & $1.57\times10^{-3}$ & $1.322\times10^{1}$ & $1.984\times10^{-21}$ \\
BP8 & 0.7 & 4.4138 & $3.681\times10^{6}$ & $1.54\times10^{-11}$ & $1.331\times10^{-3}$ & $1.236\times10^{1}$ & $1.522\times10^{-21}$ \\
BP9 & 0.7 & 5.3448 & $2.603\times10^{6}$ & $1.696\times10^{-11}$ & $2.676\times10^{-3}$ & $1.65\times10^{1}$ & $6.867\times10^{-22}$ \\
BP10 & 0.7 & 4.4138 & $3.985\times10^{5}$ & $1.658\times10^{-11}$ & $8.47\times10^{-2}$ & $7.885\times10^{1}$ & $5.082\times10^{-25}$ \\
        \hline\hline
    \end{tabular}
\end{table*}

The selected points span characteristic PBH masses from
$3.985\times10^{5}\,{\rm g}$ to $4.024\times10^{6}\,{\rm g}$.  Within the approximation adopted in
Sec.~\ref{subsec:pbh_mass_abundance}, the characteristic mass scales as
\begin{equation}
    M_\star
    \simeq
    \frac{v_{\rm eff}^2}{2G H_i}
    \left(
        e^{N_\star}-1
    \right)^2.
    \label{eq:numerical_Mstar_scaling}
\end{equation}
Consequently, variations in $v_{\rm eff}$ modify the mapping between the inflationary
nucleation history and the PBH mass scale. Since the remaining model
parameters are varied simultaneously, however, the two samples should not be
related by a simple rescaling with $v_{\rm eff}$.

For all BPs, the PBH-to-radiation ratio at horizon entry remains
small,
\begin{equation}
    \kappa_H^{\rm SB}\ll1,
\end{equation}
with the values in Table~\ref{tab:numerical_benchmarks} ranging from
approximately $1.5\times10^{-11}$ to $1.2\times10^{-8}$. The Universe is
therefore radiation dominated when the PBHs form. The points for which the
density fluctuation GW contribution is evaluated subsequently satisfy the
condition for an intermediate PBH dominated epoch before evaporation, together
with the pre-BBN evaporation requirement discussed in
Sec.~\ref{subsec:density_fluctuations}.

The same PBH population can generate a baryon asymmetry through the explosive
stage of Hawking evaporation. For the benchmark CP-violating scale
$\Lambda=1\,{\rm TeV}$, the normalized baryon yield can reach
$Y_{\rm SB}\sim{\cal O}(10^{-1})$ among the BPs presented in
Table~\ref{tab:numerical_benchmarks}. In particular, BP5 and BP10 give
\begin{equation}
    Y_{\rm SB}^{\rm BP5}
    =
    8.307\times10^{-2},
    \qquad
    Y_{\rm SB}^{\rm BP10}
    =
    8.470\times10^{-2},
    \label{eq:numerical_Y_BP5_BP10}
\end{equation}
corresponding to approximately $8\%$ of the observed baryon asymmetry for the
benchmark interaction adopted here. Moreover, BP6 yields
$Y_{\rm SB}^{\rm BP6}=6.891\times10^{-2}$ together with a density fluctuation
GW peak amplitude of $\Omega_{\rm GW,0}^{\rm DF,peak}h^2=2.998\times10^{-9}$.
These benchmarks illustrate that a sizable baryon asymmetry and a
potentially interesting GW signal can arise from the same PBH population.

As emphasized in Sec.~\ref{subsec:baryon_asymmetry}, these values should be
regarded as order-of-magnitude estimates of the baryogenesis potential of the
scenario. With the normalization adopted there, the baryon yield scales as
$Y_{\rm SB}\propto({\rm TeV}/\Lambda)^2$.
The numerical results quoted here therefore refer specifically to the
benchmark $\Lambda=1\,{\rm TeV}$ and should not be interpreted as a
model-independent upper limit on the baryon asymmetry.

The final baryon yield depends on both the integrated baryon
production and the entropy generated during PBH evaporation,
as encoded in the ratio $n_B^{\rm co}/s_{{\rm ent},\max}^{\rm co}$.
In the PBH dominated regime, increasing the initial PBH abundance
enhances both the integrated baryon production and the entropy
generated by evaporation. At sufficiently large PBH abundances,
these two effects can approximately compensate each other,
leading to a saturation of the final baryon yield, as discussed
in Ref.~\cite{Klipfel:2026nzx}. However, the BPs
presented here do not necessarily reach this saturation regime.
The numerical values of $Y_{\rm SB}$ obtained here should
therefore be regarded as representative results rather than
as an upper limit on the baryon asymmetry achievable within
the present scenario. Larger baryon yields may be obtained
for other choices of the spectator potential parameters,
subject to the relevant physical consistency conditions.
Moreover, the baryon yield remains sensitive to the details
of the baryogenesis mechanism, particularly the strength
of CP violation and the dynamics of the final PBH explosion.

\begin{figure}[!htbp]
    \centering
    \includegraphics[
        width=\textwidth
    ]{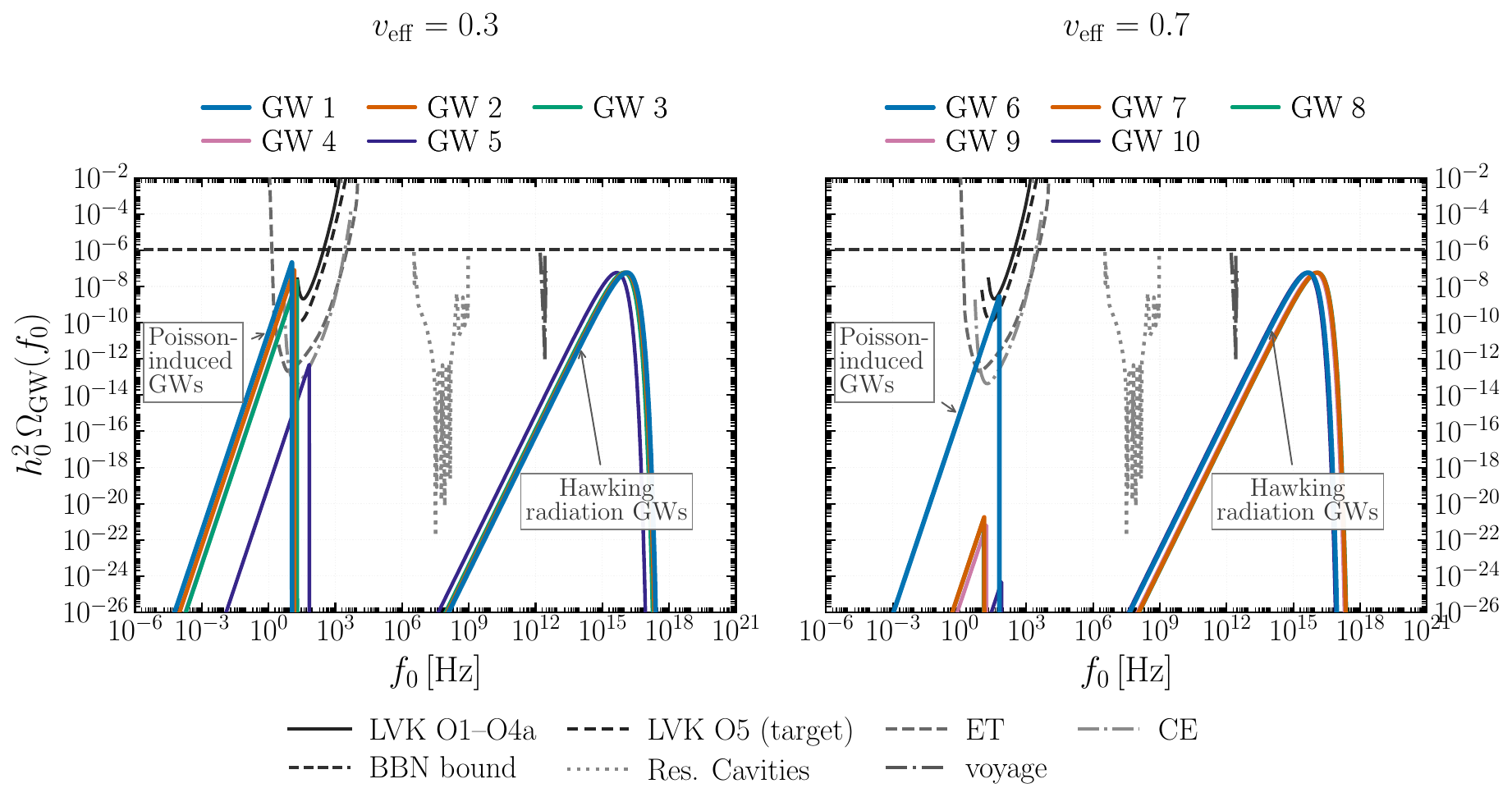}

    \caption{
    Present-day GW spectra for the BPs listed in
    Table~\ref{tab:numerical_benchmarks}.
    The curves labeled GW~1--GW~10 correspond to BP1--BP10,
    respectively. The left and right panels correspond to
    $v_{\rm eff}=0.3$ and $v_{\rm eff}=0.7$, respectively,
    with $\Delta N_{\rm nuc}=1$ in both cases.
    The low frequency and high frequency contributions originate
    from PBH density fluctuations during the PBH dominated epoch
    and direct Hawking emission of gravitons, respectively.
    For comparison, we show the PLISCs for
    LVK O1--O4a and the LVK O5 target sensitivity curve~\cite{Virgo:2025aai}, together with the projected sensitivities
    of ET~\cite{Punturo:2010zz} and CE~\cite{Reitze:2019iox}
    obtained from Ref.~\cite{Schmitz:2020syl}.
    Representative high frequency sensitivities from resonant
    electromagnetic detectors~\cite{Herman:2022fau} and searches
    based on the inverse Gertsenshtein effect~\cite{He:2023xoh}
    are also shown, along with the BBN reference bound~\cite{Aggarwal:2025noe,Caprini:2018mtu}.
    }

    \label{fig:GW_benchmarks}
\end{figure}

We now turn to the GW spectra of the BPs listed in
Table~\ref{tab:numerical_benchmarks}.
Fig.~\ref{fig:GW_benchmarks} shows the present-day spectra
for $v_{\rm eff}=0.3$ in the left panel and
$v_{\rm eff}=0.7$ in the right panel.
To assess their potential observational relevance,
we compare the predicted spectra with the
power-law-integrated sensitivity curves (PLISCs)
obtained from the combined LIGO--Virgo--KAGRA (LVK)
O1--O4a observations, together with the O5 target sensitivity curve~\cite{Virgo:2025aai}.
We also include the projected sensitivities of the
next-generation ground-based GW observatories,
the Einstein Telescope (ET)~\cite{Punturo:2010zz}
and Cosmic Explorer (CE)~\cite{Reitze:2019iox},
using the corresponding sensitivity curves provided
in Ref.~\cite{Schmitz:2020syl}.

The two well-separated features in Fig.~\ref{fig:GW_benchmarks} arise from
different production mechanisms. The lower frequency contribution is induced
by PBH density fluctuations during the intermediate PBH dominated epoch,
whereas the high frequency contribution originates from gravitons emitted
directly through Hawking evaporation. Their widely separated characteristic
frequencies therefore reflect the distinct physical scales involved in these
two mechanisms.

For $v_{\rm eff}=0.3$, the density fluctuation GW signals
of the displayed BPs have peak frequencies
in the range
\begin{equation}
    11.48\,{\rm Hz}
    \lesssim
    f_{\rm peak}^{\rm DF}
    \lesssim
    67.74\,{\rm Hz},
    \label{eq:numerical_DF_frequency_03}
\end{equation}
with peak amplitudes spanning several orders of magnitude,
\begin{equation}
    4.939\times10^{-13}
    \lesssim
    \Omega_{\rm GW,0}^{\rm DF,peak}h^2
    \lesssim
    2.26\times10^{-7}.
    \label{eq:numerical_DF_range_03}
\end{equation}
The upper end of this amplitude range is realized by BP1,
\begin{equation}
    f_{\rm peak}^{\rm DF}
    =
    11.48\,{\rm Hz},
    \qquad
    \Omega_{\rm GW,0}^{\rm DF,peak}h^2
    =
    2.26\times10^{-7},
    \label{eq:numerical_DF_BP1}
\end{equation}
while BP2--BP4 yield peak amplitudes of order $10^{-8}$.
The peak frequencies of BP1--BP4 lie in the range
of approximately $10$--$20\,{\rm Hz}$.
Despite their relatively large amplitudes, the corresponding
signals remain beyond the sensitivity of the LVK O1--O4a
observations~\cite{Virgo:2025aai}.
However, the improved sensitivity expected for LVK O5
offers better prospects for probing these BPs,
while the enhanced low frequency sensitivity of
next-generation ground-based GW observatories,
particularly ET and CE, provides further opportunities
for their detection.
In contrast, the substantially weaker GW signal
from BP5 makes its detection more challenging,
even with next-generation GW observatories.

For $v_{\rm eff}=0.7$, the displayed BPs exhibit
a comparable frequency range but a substantially wider
variation in the peak amplitude. The peak frequencies extend over
\begin{equation}
    12.36\,{\rm Hz}
    \lesssim
    f_{\rm peak}^{\rm DF}
    \lesssim
    78.85\,{\rm Hz},
    \label{eq:numerical_DF_frequency_07}
\end{equation}
with the strongest density fluctuation GW signal obtained for BP6,
\begin{equation}
    f_{\rm peak}^{\rm DF}
    =
    62.60\,{\rm Hz},
    \qquad
    \Omega_{\rm GW,0}^{\rm DF,peak}h^2
    =
    2.998\times10^{-9}.
    \label{eq:numerical_DF_BP6}
\end{equation}
Although this signal remains beyond the sensitivity of
the LVK O1--O4a observations~\cite{Virgo:2025aai},
it represents a potential target for future GW searches
with LVK O5, ET, and CE.
The same BP also yields approximately
$7\%$ of the observed baryon asymmetry for
$\Lambda=1\,{\rm TeV}$, illustrating the possibility
of probing a PBH population that contributes to
the cosmological baryon asymmetry.
In contrast, the signals from BP7--BP9 are strongly suppressed,
with peak amplitudes of order $10^{-21}$ or smaller,
while BP10 exhibits an even weaker signal with a peak
amplitude of $5.082\times10^{-25}$.
The signals from these four BPs remain far
below the projected sensitivities of the ground-based
GW experiments considered here.

The substantial variation in the peak amplitudes,
despite the comparable characteristic frequencies,
can be understood from the strong dependence of the
density fluctuation GW signal on both the PBH abundance
and mass. From Eq.~\eqref{eq:Omega_GW_density_peak},
\begin{equation}
    \Omega_{\rm GW,0}^{\rm DF,peak}
    \propto
    \left(
        \kappa_H^{\rm SB}
    \right)^{16/3}
    M_\star^{34/9}.
    \label{eq:numerical_DF_abundance_scaling}
\end{equation}
Although BP7--BP9 have comparatively large PBH masses,
their abundances are only of order $10^{-11}$,
and the steep dependence on $\kappa_H^{\rm SB}$
suppresses their density fluctuation signals by many
orders of magnitude.
For BP10, the smaller PBH mass provides an additional
suppression of the GW amplitude.
The interplay between the PBH mass and abundance
is particularly evident when comparing BP1 and BP6,
which produce the strongest signals in their respective
benchmark samples.
Although BP6 has a larger initial PBH abundance than BP1,
its smaller characteristic mass results in a lower
GW peak amplitude.

These comparisons illustrate that GW signals induced by
PBH density fluctuations can remain below the sensitivity
of current LVK observations while being potentially accessible
to future ground-based GW experiments. The improved sensitivities
expected for LVK O5, ET, and CE may therefore provide an
opportunity to probe an early PBH dominated epoch originating
from the spectator sector FOPT during inflation.
Although the predicted GW spectra are subject to
the approximations adopted in our calculations,
the comparison with the PLISCs~\cite{Thrane:2013oya}
provides a useful implication of their observational prospects.
A quantitative assessment of detectability would require
a dedicated signal-to-noise ratio analysis based on the
predicted GW spectra~\cite{Paul:2025kdd}.
Beyond assessing the detectability of these GW signals,
future investigations could also explore the possibility of
constraining or reconstructing the underlying spectator sector
model parameters from GW observations, following the recent
developments in the reconstruction of particle physics models
from cosmological phase transition GW signals~\cite{Yang:2025ybx}.
We leave these investigations for future work.

In contrast to the GWs induced by PBH density fluctuations,
the GW spectra from Hawking radiation peak at much higher
frequencies, of order $10^{16}\,{\rm Hz}$ for the benchmark
PBH masses considered here.
Their peak amplitudes vary much less across the benchmark
points than those of the induced GW spectra, the latter
being highly sensitive to the initial PBH abundance.
The main obstacle to detecting GWs from Hawking radiation
is their ultra-high peak frequencies.
Resonant electromagnetic detectors and searches based on
graviton-photon conversion have been proposed to probe
high-frequency GWs~\cite{Herman:2022fau,He:2023xoh}.
However, as shown in Fig.~\ref{fig:GW_benchmarks},
the predicted spectra from Hawking radiation remain beyond
the reach of the high frequency GW searches considered here.
The GWs induced by PBH density fluctuations therefore provide
a more promising probe of our scenario.

The comparison between the two panels also illustrates the nontrivial
dependence of the GW spectra on the underlying model parameters.
The effective wall velocity changes the bubble scale acquired during
inflation and hence the PBH mass associated with a given nucleation
history, while the spectator sector parameters simultaneously determine
the tunneling rate and the PBH abundance. Consequently, the peak frequency of the induced GW signal
is determined by $M_\star$, whereas its peak amplitude
depends on both $M_\star$ and $\kappa_H^{\rm SB}$.
The resulting GW spectra therefore reflect the interplay
between the bubble growth and nucleation histories,
rather than the value of $v_{\rm eff}$ alone.

A sizable baryon yield does not necessarily imply
a strong GW signal induced by PBH density fluctuations. For instance,
BP6 and BP10 yield comparable baryon asymmetries despite their density
fluctuation GW peak amplitudes differing by nearly sixteen orders of
magnitude. This illustrates the different dependence of baryogenesis
and induced GW production on the PBH mass and initial abundance.
While the induced GW amplitude is highly sensitive to the initial
PBH abundance, the final baryon yield is also governed by the thermal
history and entropy production during PBH evaporation.

Overall, the numerical results exhibit a characteristic two-scale GW
signature produced by the same PBH population: a very high frequency component from direct graviton
emission via Hawking radiation and a much lower frequency
component induced by PBH density fluctuations. In the parameter region
considered here, the latter extends into the frequency range relevant for
ground-based interferometers and provides the more promising observational
probe of the scenario.

\section{Conclusions}
\label{sec:conclusions}

In this work, we have investigated the formation of light
PBHs from vacuum bubbles nucleated
during a spectator sector FOPT in
inflation, together with their subsequent cosmological
consequences. A central feature of our scenario is that
the spectator potential evolves with the rolling inflaton,
allowing vacuum bubbles to nucleate and grow during inflation
before the vacuum ordering reverses near its end.
The resulting bubbles contain metastable false vacuum
interiors, whose stability against reverse vacuum tunneling
is essential for their subsequent evolution.
We account for this requirement by imposing benchmark
stability conditions before and after reheating.
Sufficiently large bubbles can then undergo supercritical
evolution and form PBHs in the parent universe~\cite{Garriga:2015fdk}.
Within this framework, the PBH abundance is determined
by the integrated vacuum tunneling history, while the
characteristic mass depends on the nucleation epoch
and the accumulated bubble growth during inflation.
Our numerical results further indicate that the PBH-producing
nucleation history can remain sharply localized even when
the spectator potential evolves over a broader interval,
supporting the monochromatic approximation adopted in
our benchmark analysis.

We have further investigated the Hawking evaporation of
the resulting PBH population and its impact on the
cosmological evolution of radiation and entropy. When the remaining PBH lifetime becomes comparable to the
LPM thermalization scale, the final stage of evaporation
can lead to an explosive release of energy into the surrounding plasma.
Following the mechanism of Ref.~\cite{Klipfel:2026nzx}, this energy deposition
can restore electroweak symmetry locally even after sphaleron processes in the
ambient Universe have frozen out. The resulting finite interval between
ambient sphaleron freeze-out and the loss of local electroweak restoration
determines the baryogenesis window in our scenario.

For the benchmark choice $\Lambda=1\,{\rm TeV}$ adopted in
Ref.~\cite{Klipfel:2026nzx}, we find that the normalized baryon yield can reach
\begin{equation}
    Y_{\rm SB}
    \equiv
    \frac{\eta_{B,\rm SB}^{\rm tot}}
         {\eta_B^{\rm obs}}
    \sim
    {\cal O}(10^{-1})
\end{equation}
for some of the BPs considered in our analysis. Thus, for
the benchmark interaction adopted here, exploding PBHs produced by the
spectator sector FOPT can generate a baryon asymmetry at the
ten-percent level relative to the observed value. At sufficiently large PBH abundance, entropy production during
evaporation can compensate for the enhancement of baryon production,
potentially leading to a saturation of the final baryon yield,
as discussed in Ref.~\cite{Klipfel:2026nzx}.
However, the numerical results presented in this work should not be
interpreted as representing this saturation limit.
Larger baryon asymmetries may be achievable in other regions of
the spectator sector parameter space or through different realizations
of the baryogenesis mechanism.

The present calculation should be interpreted as an
order-of-magnitude estimate of the baryogenesis potential
of this mechanism.
Several ingredients, including the bubble wall evolution, the mapping from the
bubble scale acquired during inflation to the PBH mass, the hydrodynamic
evolution of the final explosion, and the effective CP-violating interaction,
have been treated phenomenologically. Nevertheless, the fact that the resulting
baryon asymmetry can reach a sizable fraction of the observed value suggests
that a spectator sector FOPT followed by PBH formation and
evaporation may provide a phenomenologically relevant source of baryon number.

The required new physics enters at several stages of the mechanism. In
addition to the scalar degrees of freedom responsible for the spectator sector
FOPT, baryogenesis requires CP-violating interactions beyond those
of the Standard Model. In the present work, these effects have been
parameterized by the same effective higher-dimensional operator and the
benchmark scale $\Lambda=1\,{\rm TeV}$ used in
Ref.~\cite{Klipfel:2026nzx}. We have not attempted to construct a complete ultraviolet realization
of this operator. The nature of the CP-violating source is generally
model dependent, as illustrated by various realizations
of electroweak baryogenesis~\cite{Roy:2025zvo}. Other CP-violating operators,
as well as additional interactions between the spectator sector
and the Standard Model, could modify the baryon yield and may open
regions of parameter space in which a larger asymmetry is obtained.
Exploring these possibilities in explicit particle physics realizations
is an important direction for future work.

The same PBH population also gives rise to GW signals over widely separated
frequency ranges. Direct Hawking emission of gravitons provides a generic
high frequency contribution, while an additional signal induced by PBH density
fluctuations can arise if the PBH abundance is sufficiently large to generate
an intermediate PBH dominated epoch before evaporation~\cite{Domenech:2020ssp,Ireland:2023avg}. For the BPs considered
here, the density fluctuation contribution can reach peak amplitudes of
$\Omega_{\rm GW,0}^{\rm DF,peak}h^2\sim10^{-8}$ at frequencies of order
$10$--$10^2\,{\rm Hz}$, making it particularly interesting for ground-based
GW experiments. In contrast, the GW contribution from direct graviton
emission peaks at frequencies of order $10^{16}\,{\rm Hz}$,
posing a substantially greater challenge for
experimental detection. These two signals probe different
aspects of the PBH population and provide complementary signatures of the
inflationary spectator sector dynamics.

In the present analysis, the GW signal induced by PBH density fluctuations has
been evaluated using the monochromatic approximation. The finite duration of
the nucleation process, however, generically gives rise to a PBH mass
distribution with a nonzero width, even when the distribution is sufficiently
narrow for a monochromatic treatment to provide a useful leading
approximation. Recent work has shown that a finite-width mass distribution of
evaporating PBHs can suppress the induced GW signal relative to the idealized
monochromatic case, owing to the continuous depletion of PBHs with different
evaporation times~\cite{He:2026neq,Inomata:2020lmk,Papanikolaou:2022chm,
Gouttenoire:2026pmp}. Such a suppression could relax the corresponding GW
constraint on the PBH abundance, potentially allowing
larger initial PBH abundances. However, this does not
necessarily translate into a proportional enhancement
of the final baryon yield, since a larger PBH abundance
can also lead to more substantial entropy production
during evaporation. A quantitative assessment therefore
requires a joint treatment of the extended PBH mass
distribution, its evaporation and entropy-production
history, the baryogenesis window, and the resulting
GW spectrum. We leave such an analysis for future work.

A more complete treatment would require solving the time-dependent bubble wall
dynamics, including the evolution of the vacuum pressure throughout the
inflationary feature, together with a more precise determination of the
supercritical bubble evolution, the resulting PBH mass distribution, and its
subsequent evaporation history. On the baryogenesis side, it would be
particularly valuable to study the hydrodynamic explosion and electroweak
transport together with explicit ultraviolet completions of the CP-violating
operators. These developments would allow a more precise determination of the
PBH abundance, the final baryon asymmetry, and the associated GW signals.
Nevertheless, the present analysis illustrates how a single spectator
sector phase transition during inflation can connect vacuum
bubble nucleation, light PBH formation, Hawking evaporation, baryogenesis, and GW
production, providing a framework in which these phenomena emerge from a unified cosmological history.

\acknowledgments
This work is supported by the National Natural Science Foundation of China (NNSFC) Grant
No.~12475111 and the Fundamental
Research Funds for the Central Universities, Sun Yat-sen University.
\appendix

\section{Vacuum decay and stability at the end of inflation}
\label{app:vacuum_decay}

This appendix collects the technical details of the vacuum decay calculation
used in Sec.~\ref{sec:spectator}. We first give the explicit mapping of the
spectator potential to the semi-analytic de Sitter tunneling fit of
Ref.~\cite{Sasaki:1994yj}. We then rewrite the potential at the end of
inflation in a shifted field basis in which the nonzero metastable vacuum is
located at the origin, and use this form to estimate its decay both before and
after reheating. The post-reheating decay is treated as a zero-temperature
$O(4)$ process using the quartic-potential fit of
Ref.~\cite{Adams:1993zs}.

\subsection{CDL tunneling during inflation}
\label{appsub:sst_inflation}

At fixed $N$, the spectator potential can be written in the form used in
Ref.~\cite{Sasaki:1994yj},
\begin{equation}
    U_N(s)
    =
    \frac{1}{2}m_{\rm SST}^2s^2
    -
    \mu_{3,{\rm SST}}s^3
    +
    \lambda_{4,{\rm SST}}s^4,
\end{equation}
with
\begin{equation}
    m_{\rm SST}^2(N)
    =
    m^2(N),
    \qquad
    \mu_{3,{\rm SST}}
    =
    \frac{b}{3},
    \qquad
    \lambda_{4,{\rm SST}}
    =
    \frac{\kappa}{4}.
    \label{eq:app_SST_mapping}
\end{equation}
The corresponding dimensionless variables are
\begin{equation}
    \varepsilon_{\rm SST}(N)
    =
    \delta_N,
    \label{eq:app_epsilon_sst}
\end{equation}
\begin{equation}
    {\cal H}_{\rm SST}(N)
    =
    \frac{H_i}
         {m_{\rm deg}\sqrt{1-\delta_N}},
    \label{eq:app_H_sst}
\end{equation}
and
\begin{equation}
    \Omega_{\rm SST}(N)
    =
    \sqrt{
        \varepsilon_{\rm SST}^2(N)
        +
        {\cal H}_{\rm SST}^2(N)
    }.
    \label{eq:app_Omega_sst}
\end{equation}
The CDL tunneling exponent is then
\begin{equation}
    B_{\rm CDL}^{\rm SST}(N)
    =
    \frac{\pi^2}{6\kappa}
    (1-\delta_N)
    \widetilde B_{\rm SST}
    \left[
        \varepsilon_{\rm SST}(N),
        \Omega_{\rm SST}(N)
    \right].
    \label{eq:app_B_CDL}
\end{equation}

For completeness, the fitting function used in the numerical calculation is
\begin{align}
    \widetilde B_{\rm SST}(\varepsilon,\Omega)
    &=
    \frac{
        4
        \left(
            1+10.07\varepsilon+16.55\varepsilon^2
        \right)
    }{
        \Omega
        (\varepsilon+\Omega)^2
        (1+10\varepsilon)
    }
    \nonumber\\
    &\quad\times
    \left[
        1
        +
        \frac{11}{2}\varepsilon
        +
        \frac{3\varepsilon^2}{\Omega}
        +
        \frac{3\varepsilon^3}{2\Omega^2}
    \right]
    \widetilde B_2(\varepsilon,\Omega),
    \label{eq:app_Btilde_fit}
\end{align}
where
\begin{equation}
    \widetilde B_2(\varepsilon,\Omega)
    =
    1+
    \sum_{i,j}
    k_{ij}\varepsilon^i\Omega^j,
\end{equation}
with nonzero coefficients
\begin{align}
    k_{01}&=-0.1617,
    &
    k_{03}&=-5.507,
    \nonumber\\
    k_{11}&=-11.34,
    &
    k_{12}&=30.17,
    &
    k_{14}&=-21.69,
    \nonumber\\
    k_{20}&=12.09,
    &
    k_{21}&=-33.24,
    &
    k_{23}&=41.29,
    \nonumber\\
    k_{30}&=7.728,
    &
    k_{32}&=-19.34.
\end{align}
The characteristic tunneling exponent associated with the nucleation rate
concentrated around $N_\star$ is
\begin{equation}
    B_\star
    \equiv
    B_{\rm CDL}^{\rm SST}(N_\star).
    \label{eq:app_Bstar}
\end{equation}

Neglecting gravitational backreaction in the SST treatment requires
\begin{equation}
    \kappa_{\rm SST}(N)
    \equiv
    \frac{
        8\pi G\,m_{\rm SST}^4(N)
    }{
        \mu_{3,{\rm SST}}^2
    }
    =
    \frac{
        32\pi G b^2
    }{
        9\kappa^2
    }
    (1-\delta_N)^2.
    \label{eq:app_kappa_sst}
\end{equation}
At the characteristic nucleation point, we therefore require
\begin{equation}
    \kappa_{{\rm SST},\star}
    \ll
    \min
    \left\{
        1,\,
        \max
        \left(
            {\cal H}_{{\rm SST},\star},
            \varepsilon_{{\rm SST},\star}
        \right)
    \right\}.
    \label{eq:app_sst_backreaction}
\end{equation}

The magnitude of the curvature at the barrier top is
\begin{equation}
    \mu_{\rm top}^2(N)
    \equiv
    -U_N''(s_{-,N})
    =
    \frac{b^2}{18\kappa}
    q_N(3-q_N).
    \label{eq:app_mu_top}
\end{equation}
Within the quasi-static, weak-backreaction quartic treatment, we impose
\begin{equation}
    \frac{
        \mu_{{\rm top},\star}^2
    }{
        H_i^2
    }
    \gtrsim
    4
    \label{eq:app_HM_veto}
\end{equation}
to select the nonconstant CDL branch rather than the Hawking--Moss
regime~\cite{Sasaki:1994yj}. This condition is used as a branch-selection
criterion within the present quartic approximation and should not be
interpreted as a general criterion for an arbitrary scalar potential.

\subsection{Shifted potential at the end of inflation}
\label{appsub:endpoint_shift}

At the end of inflation, the $s=0$ configuration is the true vacuum, whereas
the nonzero minimum is a positive energy metastable state. We define
\begin{equation}
    q_{\rm end}
    \equiv
    \sqrt{1+8\delta_{\rm end}},
    \qquad
    0<q_{\rm end}<1,
    \label{eq:app_q_end}
\end{equation}
so that
\begin{equation}
    s_{+,{\rm end}}
    =
    \frac{b}{6\kappa}
    (3+q_{\rm end}).
\end{equation}
To describe tunneling from this metastable state, it is convenient to shift
the field according to
\begin{equation}
    \varphi
    \equiv
    s_{+,{\rm end}}-s.
    \label{eq:app_shift_phi}
\end{equation}
The metastable vacuum is then located at $\varphi=0$, while the true vacuum
$s=0$ corresponds to
\begin{equation}
    \varphi_{\rm tv}
    =
    s_{+,{\rm end}}
    =
    \frac{b}{6\kappa}
    (3+q_{\rm end}).
    \label{eq:app_phi_tv}
\end{equation}

Subtracting the metastable vacuum energy, we define
\begin{equation}
    \Delta U_{\rm in}(\varphi)
    \equiv
    U_{\rm end}
    \left(
        s_{+,{\rm end}}-\varphi
    \right)
    -
    U_{\rm end}
    \left(
        s_{+,{\rm end}}
    \right).
\end{equation}
Using the stationarity condition for $s_{+,{\rm end}}$, the shifted potential
takes the form
\begin{equation}
    \Delta U_{\rm in}(\varphi)
    =
    \frac{\kappa}{4}\varphi^4
    -
    \frac{b}{6}
    (1+q_{\rm end})
    \varphi^3
    +
    \frac{b^2}{36\kappa}
    q_{\rm end}(3+q_{\rm end})
    \varphi^2.
    \label{eq:app_endpoint_shifted}
\end{equation}
The barrier top in the shifted variable is
\begin{equation}
    \varphi_{\rm top}
    =
    s_{+,{\rm end}}
    -
    s_{-,{\rm end}}
    =
    \frac{b}{3\kappa}
    q_{\rm end}.
    \label{eq:app_endpoint_top}
\end{equation}
Eq.~\eqref{eq:app_endpoint_shifted} is the potential used below to
estimate the lifetime of the metastable bubble interior.

\subsection{Vacuum decay before reheating}
\label{appsub:endpoint_dS}

Before instantaneous reheating, the metastable state present at the end of
inflation is embedded in the inflationary de Sitter background. Its decay can
therefore be estimated using the same semi-analytic de Sitter tunneling
description introduced above, provided that the
corresponding nonconstant bounce solution exists.

Matching Eq.~\eqref{eq:app_endpoint_shifted} to
\begin{equation}
    \Delta U_{\rm in}(\varphi)
    =
    \frac{1}{2}
    m_{{\rm SST,end}}^2\varphi^2
    -
    \mu_{3,{\rm end}}\varphi^3
    +
    \lambda_{4,{\rm end}}\varphi^4
\end{equation}
gives
\begin{equation}
    m_{{\rm SST,end}}^2
    =
    \frac{b^2}{18\kappa}
    q_{\rm end}(3+q_{\rm end}),
    \label{eq:app_m_SST_end}
\end{equation}
and
\begin{equation}
    \mu_{3,{\rm end}}
    =
    \frac{b}{6}
    (1+q_{\rm end}),
    \qquad
    \lambda_{4,{\rm end}}
    =
    \frac{\kappa}{4}.
    \label{eq:app_endpoint_SST_coefficients}
\end{equation}
The corresponding SST variables are
\begin{equation}
    \varepsilon_{{\rm SST,end}}
    =
    \frac{
        1-q_{\rm end}
    }{
        (1+q_{\rm end})^2
    },
    \label{eq:app_endpoint_epsilon}
\end{equation}
\begin{equation}
    {\cal H}_{{\rm SST,end}}
    =
    \frac{
        H_i
    }{
        m_{{\rm SST,end}}
    },
    \qquad
    \Omega_{{\rm SST,end}}
    =
    \sqrt{
        \varepsilon_{{\rm SST,end}}^2
        +
        {\cal H}_{{\rm SST,end}}^2
    }.
    \label{eq:app_endpoint_SST_variables}
\end{equation}
When the nonconstant CDL branch is applicable, the corresponding exponent is
\begin{equation}
    B_{\rm CDL,end}^{\rm SST}
    =
    \frac{\pi^2}{6\kappa}
    \frac{
        q_{\rm end}(3+q_{\rm end})
    }{
        (1+q_{\rm end})^2
    }
    \widetilde B_{\rm SST}
    \left(
        \varepsilon_{{\rm SST,end}},
        \Omega_{{\rm SST,end}}
    \right).
    \label{eq:app_B_CDL_end}
\end{equation}

The gravitational-backreaction parameter associated with the potential at the
end of inflation is
\begin{equation}
    \kappa_{{\rm SST,end}}
    \equiv
    \frac{
        8\pi G
        m_{{\rm SST,end}}^4
    }{
        \mu_{3,{\rm end}}^2
    }
    =
    \frac{
        8\pi G b^2
    }{
        9\kappa^2
    }
    \frac{
        q_{\rm end}^2
        (3+q_{\rm end})^2
    }{
        (1+q_{\rm end})^2
    }.
    \label{eq:app_kappa_SST_end}
\end{equation}
Accordingly, Eq.~\eqref{eq:app_B_CDL_end} is used only when
\begin{equation}
    \kappa_{{\rm SST,end}}
    \ll
    \min
    \left\{
        1,\,
        \max
        \left(
            {\cal H}_{{\rm SST,end}},
            \varepsilon_{{\rm SST,end}}
        \right)
    \right\}.
    \label{eq:app_endpoint_SST_backreaction}
\end{equation}

For a sufficiently broad barrier, the relevant de Sitter transition is instead
described by the Hawking--Moss saddle. From
Eq.~\eqref{eq:app_endpoint_shifted}, the barrier height above the metastable
vacuum is
\begin{equation}
    \Delta U_{\rm top,end}
    =
    \frac{b^4}{324\kappa^3}
    q_{\rm end}^3.
    \label{eq:app_DeltaU_top_end}
\end{equation}
In the spectator-subdominant limit,
\begin{equation}
    \Delta U_{\rm top,end}
    \ll
    3M_{\rm Pl}^2H_i^2,
\end{equation}
the Hawking--Moss action difference can be expanded as
\begin{equation}
    B_{\rm HM,end}
    \simeq
    \frac{
        8\pi^2
    }{
        3H_i^4
    }
    \Delta U_{\rm top,end},
\end{equation}
which gives
\begin{equation}
    B_{\rm HM,end}
    \simeq
    \frac{
        2\pi^2
    }{
        243
    }
    \frac{
        b^4
    }{
        \kappa^3H_i^4
    }
    q_{\rm end}^3.
    \label{eq:app_B_HM_end}
\end{equation}

The magnitude of the curvature at the barrier top at the end of inflation is
\begin{equation}
    \mu_{\rm top,end}^2
    =
    \frac{b^2}{18\kappa}
    q_{\rm end}
    (3-q_{\rm end}).
    \label{eq:app_mu_top_end}
\end{equation}
For the benchmark treatment adopted here, we use
\begin{equation}
    B_{\rm end}^{\rm dS}
    =
    \begin{cases}
        B_{\rm CDL,end}^{\rm SST},
        &
        \displaystyle
        \frac{\mu_{\rm top,end}^2}{H_i^2}>4,
        \\[7pt]
        B_{\rm HM,end},
        &
        \displaystyle
        \frac{\mu_{\rm top,end}^2}{H_i^2}\leq4.
    \end{cases}
    \label{eq:app_B_end_dS}
\end{equation}
The CDL expression is used only when
Eq.~\eqref{eq:app_endpoint_SST_backreaction} is also satisfied. Near the
boundary $\mu_{\rm top,end}^2/H_i^2\simeq4$, the piecewise prescription should
be regarded as an approximate branch selection; a numerical Euclidean
calculation would be required for a more precise treatment.

We require the physically relevant de Sitter transition at the end of inflation
to be exponentially more suppressed than the bubble nucleation process
responsible for the PBH source,
\begin{equation}
    B_{\rm end}^{\rm dS}
    >
    B_\star
    +
    \Delta B_{\rm safe},
    \qquad
    \Delta B_{\rm safe}\gg1.
    \label{eq:app_endpoint_dS_stability}
\end{equation}
This is an exponent-level benchmark condition rather than a calculation of the
full integrated decay probability during the remaining inflationary evolution.

\subsection{Post-reheating zero temperature bounce}
\label{appsub:endpoint_flat}

After reheating, we assume that the spectator sector remains sufficiently
decoupled from the ambient radiation bath that thermal corrections to
Eq.~\eqref{eq:app_endpoint_shifted} are negligible. The decay of the
metastable bubble interior can then be treated as a zero-temperature $O(4)$
tunneling process.

To apply the quartic-potential fit of Ref.~\cite{Adams:1993zs}, we write
\begin{equation}
    V_A(\varphi)
    =
    \lambda_A\varphi^4
    -
    a_A\varphi^3
    +
    b_A\varphi^2.
\end{equation}
Comparison with Eq.~\eqref{eq:app_endpoint_shifted} gives
\begin{equation}
    \lambda_A
    =
    \frac{\kappa}{4},
    \qquad
    a_A
    =
    \frac{b}{6}
    (1+q_{\rm end}),
\end{equation}
and
\begin{equation}
    b_A
    =
    \frac{b^2}{36\kappa}
    q_{\rm end}(3+q_{\rm end}).
\end{equation}
The corresponding shape parameter is
\begin{equation}
    D_{\rm A,end}
    \equiv
    \frac{
        8\lambda_A b_A
    }{
        a_A^2
    }
    =
    \frac{
        2q_{\rm end}(3+q_{\rm end})
    }{
        (1+q_{\rm end})^2
    }.
    \label{eq:app_D_Adams}
\end{equation}
Since $0<q_{\rm end}<1$, one has
\begin{equation}
    0
    <
    D_{\rm A,end}
    <
    2,
\end{equation}
with $D_{\rm A,end}\rightarrow2$ corresponding to the nearly degenerate
thin-wall limit.

The zero-temperature bounce exponent is then
\begin{equation}
    B_{4,\rm end}^{\rm flat}
    =
    \frac{
        4\pi^2
    }{
        3\kappa
    }
    (2-D_{\rm A,end})^{-3}
    \left(
        13.832D_{\rm A,end}
        -
        10.819D_{\rm A,end}^2
        +
        2.0765D_{\rm A,end}^3
    \right).
    \label{eq:app_B4_Adams}
\end{equation}

The post-reheating decay of the metastable bubble interior takes place in a
time-dependent cosmological background, so that the corresponding
semiclassical tunneling action need not coincide with its flat-space value.
As discussed in Ref.~\cite{Simon:2009nb}, the importance of the background
evolution can be assessed by comparing the characteristic scale of the
tunneling process with the dynamical scales of the FLRW geometry. When the
tunneling scale is much shorter than both the Hubble time and the time scale
over which the expansion rate changes, the Minkowski-space result provides a
good approximation. If these scales become comparable, the tunneling process
can become sensitive to the time dependence of the background and the
corresponding action can receive appreciable corrections. In particular,
sufficiently large bubbles can retain sensitivity to the preceding evolution
of the scale factor, including details of the reheating history~\cite{Simon:2009nb}.

The characteristic length entering this comparison depends on the structure
of the bounce. In the thick-wall regime, there is no parametrically large
separation between the microscopic wall scale and the characteristic size of
the bounce configuration, so that the inverse curvature scale at the barrier
top provides a useful order-of-magnitude proxy~\cite{Megevand:2023nin,Widyan:1998rx,Hayashi:2020ocn}. Toward the thin-wall
regime, however, the nucleation radius can become parametrically larger than
this local microscopic scale. Consequently, a condition based only on the
barrier curvature does not by itself establish the validity of the
flat-space approximation for a near-thin-wall bounce.

In the present phenomenological analysis, we nevertheless impose a minimal
local consistency requirement based on the microscopic curvature scale,
\begin{equation}
    R_{b,\rm end}
    \sim
    \left[
        -\Delta U_{\rm in}''
        \left(\varphi_{\rm top}\right)
    \right]^{-1/2}
    =
    \left[
        \frac{b^2}{18\kappa}
        q_{\rm end}(3-q_{\rm end})
    \right]^{-1/2},
    \label{eq:app_Rb_estimate}
\end{equation}
and require
\begin{equation}
    H_iR_{b,\rm end}\ll1.
    \label{eq:app_flat_validity}
\end{equation}
This condition ensures that the microscopic scale associated with the barrier
is well separated from the cosmological scale immediately after reheating,
but it does not exclude configurations whose full nucleation radius becomes
larger in the thin-wall regime. We therefore retain such configurations within
the present benchmark treatment, while interpreting the corresponding
flat-space decay estimate with this limitation in mind. A definitive
calculation in this regime would require the detailed reheating history and a
solution of the tunneling problem in the resulting time-dependent background,
which we leave for future work.

Finally, we require the zero-temperature vacuum decay after reheating to be
exponentially more suppressed than the vacuum tunneling process responsible
for bubble production,
\begin{equation}
    B_{4,\rm end}^{\rm flat}
    >
    B_\star
    +
    \Delta B_{\rm safe},
    \qquad
    \Delta B_{\rm safe}\gg1.
    \label{eq:app_endpoint_flat_stability}
\end{equation}
Together,
Eqs.~\eqref{eq:app_endpoint_dS_stability} and
\eqref{eq:app_endpoint_flat_stability} provide the benchmark stability
conditions adopted in our analysis for the metastable vacuum present at the
end of inflation. These conditions are intended as conservative criteria at
the level of the tunneling exponent and do not replace a fully time-dependent
calculation of the integrated decay probability.

\section{Spectator potential parameters for the BPs}
\label{app:benchmark_parameters}

For completeness, we collect in
Table~\ref{tab:benchmark_potential_parameters} the spectator potential
parameters corresponding to the BPs presented in
Table~\ref{tab:numerical_benchmarks}. These quantities specify the microscopic
spectator sector realization underlying each benchmark, while the PBH,
baryogenesis, and GW quantities reported in
Table~\ref{tab:numerical_benchmarks} are derived from the subsequent tunneling,
bubble, and cosmological evolution.

The precise value of $\sigma_N$ is tied to the detailed time-dependent bubble wall
dynamics, since it controls the duration over which the relative vacuum energies
evolve and therefore affects both the acceleration and the subsequent
deceleration of the wall. A self-consistent determination of $\sigma_N$ would
require solving the wall equation of motion throughout the interval over which
the spectator potential evolves.

The value of $\sigma_N$ can also influence the width of the bubble population
in scale space. A broader feature in the spectator potential generally allows
vacuum tunneling to remain relevant over a wider interval in $N$, which can in
turn broaden the range $\Delta N_{\rm PBH}$ entering the PBH mass distribution
and therefore affect the accuracy of the monochromatic approximation.
Nevertheless, the relation is not one-to-one, since the nucleation rate depends
exponentially on the bounce action,
$\Gamma(N)\propto \exp[-B_{\rm CDL}(N)]$. Consequently, a smooth potential
feature with a finite width can still generate a much more sharply localized
nucleation history when the bounce action develops a sufficiently pronounced
minimum, as explicitly illustrated in Ref.~\cite{Wang:2025hwc}. The validity of
the monochromatic approximation is therefore controlled directly by the
resulting nucleation interval $\Delta N_{\rm PBH}$ rather than by $\sigma_N$
alone.

The requirement of a small residual wall velocity at the end of
inflation may likewise select a non-negligible duration for the wall evolution.
In particular, a sufficiently extended post-reversal stage can allow the
outward motion acquired near nucleation to be gradually reduced before
the end of inflation, rather than requiring an abrupt stopping of the wall.
Such a possibility is compatible with a sizable accumulated propagation and
hence with a non-negligible $v_{\rm eff}$.

Importantly, the spectator potential contains additional freedom to adjust the
relative strength of the driving and decelerating stages. For fixed
$N_\star$ and $\sigma_N$, the vacuum structure near the nucleation epoch and at
the end of inflation is characterized by
\begin{equation}
\delta_\star
=
1-\alpha_0+\delta_m,
\qquad
\delta_{\rm end}
=
1-\alpha_0
+
\delta_m
\exp\left[
-\frac{N_\star^2}{2\sigma_N^2}
\right].
\end{equation}
The independent parameters $\alpha_0$ and $\delta_m$ therefore provide freedom
to vary the vacuum energy splitting around nucleation and the residual
metastable vacuum energy near the end of inflation separately, subject to the
vacuum-structure and stability conditions imposed in the main text. In
particular, one can favor a larger vacuum energy difference during the
outward-driven stage, enhancing the early wall propagation, while keeping the
post-reversal interior vacuum energy relatively small. The latter corresponds
to a weaker inward vacuum-pressure contribution and hence to a more gradual
deceleration. Together with the finite duration of the post-reversal stage and
the other contributions to the wall dynamics, this provides a plausible
parameter regime in which the wall can accumulate a sizable displacement during
inflation while its residual velocity becomes small by
$t_{\rm end}$. The additional parameters controlling the overall scale and
shape of the spectator potential provide further flexibility, although not
every choice of $\sigma_N$ can necessarily be compensated while simultaneously
satisfying all tunneling, stability, spectator-subdominance, and
supercriticality requirements.

In the present phenomenological treatment, we therefore fix
$\Delta N_{\rm nuc}=1$ as a common benchmark and retain parameter sets for which
viable solutions are obtained for multiple values of $\sigma_N$. This procedure
selects regions of parameter space whose phenomenological viability is less
sensitive to the precise value of $\sigma_N$, thereby reducing the dependence
of our results on the unresolved wall dynamics. We therefore do not list a
unique value of $\sigma_N$ for each BP in
Table~\ref{tab:benchmark_potential_parameters}. A dedicated calculation of the
time-dependent bubble wall evolution, together with the resulting nucleation
width and PBH mass distribution, is left for future work.

\begin{table*}[t]
    \centering
    \small
    \setlength{\tabcolsep}{6pt}
    \renewcommand{\arraystretch}{1.15}
\caption{
Spectator potential parameters corresponding to the BPs listed in
Table~\ref{tab:numerical_benchmarks}. The first five points correspond to
$v_{\rm eff}=0.3$, while the remaining five correspond to
$v_{\rm eff}=0.7$. Since the precise value of $\sigma_N$ depends on the
assumed bubble wall dynamics, we retain parameter sets that remain viable for
multiple values of $\sigma_N$ and do not list it separately.
}
    \label{tab:benchmark_potential_parameters}
\begin{tabular}{c c c c c c c}
    \hline\hline
    BP
    &
    $v_{\rm eff}$
    &
    $H_i\,[\mathrm{GeV}]$
    &
    $b\,[\mathrm{GeV}]$
    &
    $\kappa$
    &
    $\alpha_0$
    &
    $\delta_m$
    \\
    \hline
BP1 & 0.3 & $1\times10^{13}$ & $1.189\times10^{15}$ & 0.4171 & 1.0689 & 0.9425 \\
BP2 & 0.3 & $1.083\times10^{12}$ & $1.287\times10^{14}$ & 0.4171 & 1.0392 & 0.9128 \\
BP3 & 0.3 & $1\times10^{13}$ & $1.189\times10^{15}$ & 0.4171 & 1.0434 & 0.9171 \\
BP4 & 0.3 & $1.172\times10^{11}$ & $1.34\times10^{14}$ & 0.4171 & 1.0519 & 0.9255 \\
BP5 & 0.3 & $1.083\times10^{12}$ & $1.533\times10^{15}$ & 1.8 & 1.0561 & 0.7007 \\
        \hline
BP6 & 0.7 & $1\times10^{13}$ & $9.457\times10^{13}$ & 0.0967 & 1.0095 & 0.9595 \\
BP7 & 0.7 & $1\times10^{13}$ & $9.097\times10^{14}$ & 0.0967 & 1.0986 & 1.0486 \\
BP8 & 0.7 & $1.172\times10^{11}$ & $1.026\times10^{14}$ & 0.0967 & 1.0264 & 0.9764 \\
BP9 & 0.7 & $1.083\times10^{12}$ & $9.848\times10^{13}$ & 0.0967 & 1.1198 & 1.0698 \\
BP10 & 0.7 & $1.083\times10^{12}$ & $9.848\times10^{13}$ & 0.0967 & 1.124 & 1.074 \\
        \hline\hline
    \end{tabular}
\end{table*}

\end{CJK}

\newpage

\bibliographystyle{JHEP}
\bibliography{Baryogenesis}

\end{document}